\documentclass[twocolumn,showpacs,amsmath,amssymb,superscriptaddress]{revtex4-2}
\usepackage{graphicx} 

\usepackage[caption=false, listofformat=subsimple, labelformat=simple]{subfig}

\begin{document}
	
\title{Local tests of Schwarzschild-Bertotti-Robinson spacetime}
\author{Fatma Kutlu}
\affiliation{Physics Department, Marmara University, Faculty of Science, 34722 \.Istanbul, T\"urkiye}

\author{Livanur Çelik}
\affiliation{Physics Department, Marmara University, Faculty of Science, 34722 \.Istanbul, T\"urkiye}

\author{\"Ozg\"ur Delice}
\email{ozgur.delice@marmara.edu.tr}
\affiliation{Physics Department, Marmara University, Faculty of Science, 34722 \.Istanbul, T\"urkiye}

\begin{abstract}
	We study weak field local tests of Schwarzschild-Bertotti-Robinson spacetime, which describes a static black hole embedded in an external magnetic field. Perihelion precession, deflection of light rays and time delay of radar echoes are the phenomena considered in this work to understand the possible effects of this external magnetic field on these observationally important tests. We not only derive weak field analytical results by employing appropriate methods, but also consider an exact approach  to directly integrate the relevant geodesic equations, to both confirm those weak field analytical results and to obtain results beyond the weak field approximation.
\end{abstract}
\maketitle

\section{Introduction}

Black holes are fascinating objects in the cosmos with interesting properties and behavior. Before the advent of modern observational astronomy, the focus was on understanding them and the phenomena related to their behavior on theoretical grounds. A lot of effort was made to understand properties of these objects as solutions of Einstein's equations, their interaction  with surrounding test particles or other fields, their quantum mechanical or statistical insight relevant to unifying physical theories or obtaining quantum gravity theory.  In recent years, thanks to recent technological developments in  observational astronomy and astrophysics leading to new accumulated data and analysis techniques, they are observationally verified to exist. The gravitational wave observations due to merging of black holes confirmed their existence and their several properties \cite{LIGO2016}. Tracking stars in the galactic centers with high sensitivity \cite{GRAVITY2018,Ghez2019} is another method confirming their existence. The observations of phenomena around supermassive black holes in the centers of galaxies lead to imaging of such black holes \cite{EHT_M87,EHT_SgrA}. All these developments bring black holes as primary elements of observational astrophysics rather than some theoretical tool of theories of gravity.

The black holes are not just isolated objects; their astrophysical relevance usually depends on their interaction with surrounding matter and/or field distributions. For example, enormous magnetic fields can arise in some astrophysical processes of moving charged particle distributions. Hence, the interaction of black holes with these external magnetic fields can be important to analyze behavior of these matter distributions, especially the accretion disks around supermassive black holes.  In this manner, several approaches can be taken to study such interaction. A test field approach is usually enough to understand this interaction for the usual case where the magnetic field is weak such that it does not modify the background spacetime geometry \cite{Wald1974, Prasanna1980}. Hence in this approach, the effect of the magnetic field is only felt by charged particles moving around the black hole. This approach covers most of the relevant astrophysical phenomena and has been developed and applied successfully to many interesting cases \cite{Aliev1989, Frolov2010, Kolos2023_parabolic, Zelenka2026, Haddad2025}.

If the magnetic field is much stronger, it can affect the background curvature and therefore the motion of neutral test particles. To study these effects beyond test field approach, approximate or exact solutions may be needed. These require solving coupled Einstein-Maxwell equations  in the presence of black holes with nontrivial magnetic fields.  This is rather a difficult problem, hence the relevant solutions are usually obtained by some solution generating technique. One such solution \cite{Ernst1976}, called the Schwarzschild-Melvin (or Ernst) spacetime, can be obtained by applying the Harrison transformation \cite{Harrison:1968wue} to  the Schwarzschild seed. The resulting solution describes a static black hole embedded in a uniform magnetic field, which reduces to the static cylindrically symmetric Bonnor-Melvin magnetic universe \cite{Bonnor, Melvin}, when the black hole vanishes. This method was later applied to generate more general magnetized black hole solutions \cite{Ernst:1976bsr}. These magnetized solutions helped to understand the interaction of black holes with strong magnetic fields and behavior of test particles around magnetized black hole spacetimes \cite{Konoplya2006_particles,Konoplya2006_lens}. It turns out that these magnetized solutions have some drawbacks. They are type $I$ solutions \cite{BoseEsteban} where the magnetic field extends along  the whole $z$ axis and vanishes  far from black hole but the test particles cannot escape to infinity, and their motion can be chaotic \cite{Karasvok}. Therefore, these solutions  cannot describe an isolated compact magnetic field surrounding a black hole.

An isolated source in general relativity is  described by looking at its algebraic structure using relevant techniques such as Newman-Penrose method. If a solution is type $D$ in Petrov classification, then it is considered as describing compact isolated source. Obtaining general type vacuum or electrovacuum type $D$ solutions in GR has a long history. Recently  a general type $D$  Einstein-Maxwell solution describing a rotating black hole with a nonzero magnetic field was presented \cite{Podolsky2025}. The solution can be called  Kerr-Bertotti-Robinson spacetime since when the magnetic field vanishes it reduces to the Kerr solution whereas when the mass and rotation parameter vanish it reduces to Bertotti-Robinson spacetime  \cite{Bertotti:1959pf,Robinson:1959ev}. It might be argued that this particular solution  describes a black hole embedded in an external magnetic field better than Kerr-Melvin spacetime due to the following properties: this solution is type $D$,  it has bounded ergoregions and magnetic field is asymptotically uniform and finite. This solution belongs to  a large family of solutions having nonaligned principal null directions \cite{Ovcharenko:2025cpm}.   For the present work we consider a particular subcase, i.e., the nonrotating case,  which may be called  Schwarzschild-Bertotti-Robinson (SBR) black hole. Note that this SBR solution was already contained in the recent type $D$ Einstein-Maxwell solutions presented in \cite{VanDenBergh2020}, although there is an earlier work \cite{Alekseev:1996fq} with similar construction  but whether they describe same spacetime is still an open question.   Several important mathematical aspects and physical properties of the Kerr-BR solution were studied in various works  \cite{Hu:2026slp,Zeng:2025olq,Wang:2025vsx,Li:2025rtf,Wang:2025bjf,Vachher:2025jsq,Zhang:2025ole,Gray:2025lwy}.

These appealing features of Schwarzschild-BR solutions motivated us to consider their physical properties, especially weak field classical tests of GR \cite{Will:2014kxa}, with the hope of better understanding the effects of the external magnetic field on these phenomena. These classical tests have well-established  experimental or observational results. It could be interesting to obtain analytical results for the effects of this magnetic field  as  corrections to the renowned expressions for Schwarzschild black holes in the weak field regime. Moreover, these results can help to set limits on the magnetic field strength based on the solar system tests. We  also study  strong field behavior by  integrating relevant geodesic equations using a numerical approach. This will help us to see how this  magnetic field affects these phenomena in the strong field regime. 

 The paper is organized as follows:   we review the SBR metric in Section II. We study perihelion precession phenomena in Section III where the weak field analytical  and numerical approaches will be discussed as subsections. We consider deflection of light rays in Section IV and time delay of radar echoes in Section V. In all cases, we have subsections devoted to weak field analytical approach, and exact and numerical approaches for these phenomena. We finish this paper with a brief conclusion.  

\section{The Schwarzschild-Bertotti-Robinson Spacetime and its geodesics}

The Schwarzschild-Bertotti-Robinson spacetime can be obtained from Kerr-BR solution  \cite{Podolsky2025}, a recently introduced novel type $D$ solution, by setting the rotation parameter to zero. Its line element is written in the standard spherical coordinates with standard coordinate ranges as
\begin{equation}\label{SBRmnetric}
	ds^2= \frac{1}{\Omega^2}\left[ -\frac{Q}{r^2} dt^2+\frac{r^2}{Q}dr^2 +r^2 \left( \frac{d\theta^2}{P}+P \sin^2\theta d\phi^2\right) \right],\ \
\end{equation}
where the metric functions $P, Q$ and $\Omega$ are defined as
\begin{eqnarray}
	&&P=1+B^2 M^2 \cos^2\theta,  \label{SBRmetricfunc}  \\
	&& Q=\left( 1+B^2r^2\right) \Delta,  \nonumber \\
	&& \Omega^2=\left( 1+B^2 r^2\right) -B^2 \Delta \cos^2 \theta, \nonumber \\
	&&\Delta= \left( 1-B^2 M^2\right ) r^2-2M\, r. \nonumber
\end{eqnarray}
Here the constants $M$ and $B$ are related to the mass of the black hole and the external magnetic field, respectively. This line element and the vector potential given below satisfy Einstein-Maxwell field equations:
\begin{eqnarray}
	A&=&A_\phi d\phi,\\
	 A_\phi&=&\frac{\Omega-r\,\Omega_{,r}-1}{B}=\frac{1}{B} \left( \frac{1+B^2 M\, r \cos^2\theta}{\Omega}-1\right).\
\end{eqnarray}
Note that when $B=0$ the resulting line element reduces to the Schwarzschild black hole solution and when $M=0$ the solution is Bertotti-Robinson spacetime.

This black hole solution admits a single horizon given by
\begin{equation}
	r_h=\frac{2M}{1-B^2M^2},
\end{equation}
where, unlike the Ernst solution, it depends on  the magnetic field as well as the black hole mass and increases with the magnetic field strength.  The condition $B^2M^2<1$ is necessary to have a regular horizon, and  we use this condition throughout this work.

In the SBR spacetime described by equations (\ref{SBRmnetric},\ref{SBRmetricfunc}), the axis is not regular in general \cite{Podolsky2025}: for $\phi \in[0,2\pi)$ there is a conical excess, since the ratio of circumference to proper radius  for a small circle around the axis is $2\pi  (1+B^2M^2)$ instead of $2\pi$.  One can regularize this spacetime by restricting the  angular range to $\phi \in [0,2\pi/(1+B^2M^2) )$ or, equivalently, multiplying the $g_{\phi\phi}$ term by $1/(1+B^2M^2)^2$, while keeping standard angular range. Although the SBR solution and its regularized versions are definitely different solutions globally, their local physics is the same and  the angular excess only has  global effects associated with the angular identification. Therefore, we use the standard range in the following discussion but we comment on our results whenever a notable difference occurs due to regularization.

\subsection{Geodesics of SBR Black Hole}
Test particle motion in the spacetime (\ref{SBRmnetric}) can be obtained from the Lagrangian
\begin{equation}
	\mathcal{L} =\frac{1}{2} g_{\mu\nu} \dot{x}^\mu \dot{x}^\nu
\end{equation}
where overdot represents an affine parameter, $\lambda$, and for massive particles it can be taken as the proper time. Here we have normalized the Lagrangian such that the first integrals will also be normalized per unit test particle mass. Euler-Lagrange equations 
\begin{equation}\label{lagrangeeq}
	\frac{d}{d\lambda} \left( \frac{\partial \mathcal{L}}{ \partial\dot{x}^\mu }   \right)-\frac{\partial\mathcal L}{\partial x^\mu} =0,
\end{equation}
yield the following first integrals corresponding to the Killing vectors $\partial_t$ and $\partial_\phi$, respectively:
\begin{eqnarray}
	&&	E=\frac{Q}{\Omega^2 r^2}\dot{t}, \label{tdot}\\
	&&	L=\frac{P r^2 \sin^2\theta}{\Omega^2}\dot{\phi}, \label{phidot}
\end{eqnarray}
where $E$ and  $L$ are the energy and angular momentum per unit mass for massive test particles. Geodesic equations for $r$ and $\theta$ of (\ref{lagrangeeq}) are complicated, so we do not present them explicitly, as we do not need them in that form in this paper.
    
If we consider equatorial geodesics by choosing $\theta=\pi/2$, then using the line element itself by considering $ds^2/d\lambda^2=\epsilon$ with $\epsilon=-1,0$ for massive or massless test particles, the radial geodesics can be derived from the following effective potential equation
\begin{eqnarray}
	&&\dot{r}^2=(1+B^2r^2)^2\left[E^2-V(r)\right],  \label{effpot} \\
	&& V(r)=\left( 1-B^2M^2 -\frac{2M}{r}\right)\left[\left(1+B^2 r^2\right)\frac{L^2}{r^2}-\epsilon\right]. \nonumber
\end{eqnarray}
In the following, we will mainly use this effective potential since we discuss equatorial orbits for both massive and massless test particles.

\section{Timelike Geodesics, Circular geodesics and Advance of Perihelion}  

\begin{figure}
	\centering
	\includegraphics[draft=false,width=1\linewidth]{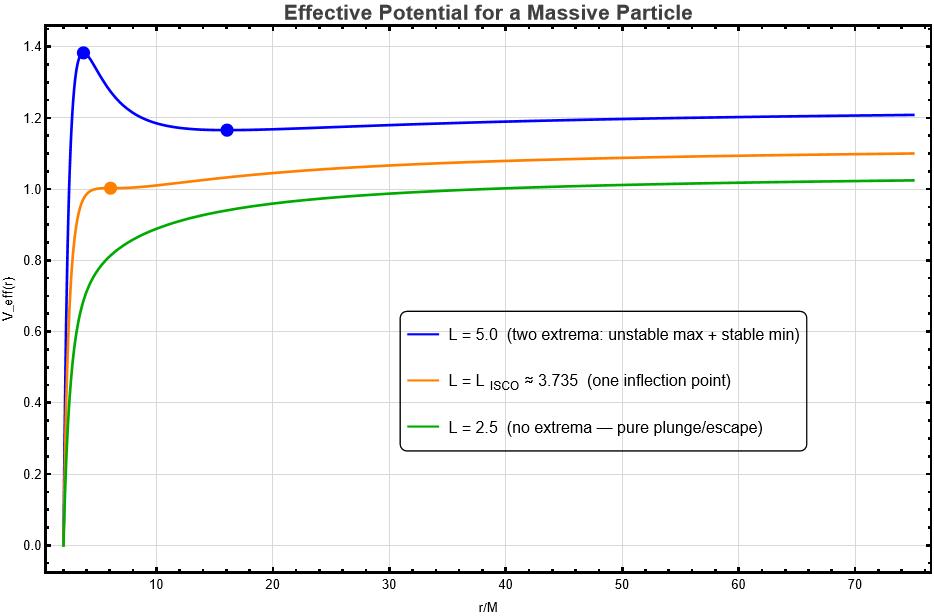}
	\caption{Effective potential for three distinct characteristic cases: two circular orbits (blue), single circular orbit (red) and none (green). The blue curve was plotted for $L=5.0$, red for $L=L_{ISCO}=3.735$, and  green for $L=2.5$. All curves are plotted for $M=1$, $B=0.1$. }
	\label{fig:effpotm}
\end{figure}

Here we consider behavior of the massive test particles  around this black hole mostly for bounded orbits. In order to investigate this motion, especially perihelion precession  of test particles (planets, etc.)  in  bounded orbits around this black hole, we need to use timelike geodesics, i. e., we take $\epsilon=-1$ in (\ref{effpot}). We also consider equatorial geodesics by taking $\theta=\pi/2$ and the geodesic equations reduce to 
\begin{align} 
	&E=\frac{\Delta}{r^2}\dot{t}, \quad L=\frac{r^2}{1+B^2r^2} \dot{\phi},\label{geo1}\\
	&\left(\frac{dr}{d\tau}\right)^2=(1+B^2r^2)^2\left[E^2-V_m(r) \right], \label{geo2}\\
	& V_m(r)=\left( 1-B^2M^2 -\frac{2M}{r}\right)\left[\left(\frac{1}{r^2}+B^2\right)L^2+1\right], \label{effpotm}
\end{align}  
where now overdot represents proper time $\tau$ of the massive test particle.  Here $V_m$ represents the effective potential of the massive test particles.  The generic behavior of the effective potential (\ref{effpotm}) can be seen in the Fig.~\ref{fig:effpotm}, in which, depending on the parameters of the black hole and the test particle, there could be two extremum points, single extremum point or none. Some aspects of equatorial geodesics of SBR geometry have already been discussed,   such as  periodic bound orbits, gravitational waveforms \cite{Xamidov:2026kqs}, and orbital and epicyclic  motion of charged particles \cite{Xamidov:2026kqs,Xu:2026tgb}. Therefore we consider some other aspects not yet discussed in these previous works.

\subsection{Circular Geodesics}

We can determine circular orbits for massive test particles by the condition $\dot{r}=\ddot{r}=0$, which requires $V_m(r_{cm})=E^2$, $dV_m/dr=0$.
The latter condition for massive test particles gives 
\begin{eqnarray}\label{rcircmassive}
	r^{\pm}_{ cm}&=& \frac{1}{2M(1+B^2L^2)} \bigg[ L^2 \left(1-B^2 M^2\right) \\
&& \pm\,  L\sqrt{L^2(1-B^2M^2) -12 M^2 (1+B^2L^2)} \bigg],\ \  \nonumber 
\end{eqnarray}
which reduce to corresponding Schwarzschild values when $B$ vanishes. In general, test particles in this spacetime can follow circular geodesics at two radii for given black hole parameters and energy and angular momentum of the test particle. The inner circular orbit at $r^-_{cm}$ is unstable, whereas the outer one  at $r^+_{cm}$ is stable, which can be seen by evaluating the second derivative of effective potential  at these radii.

The radius, angular momentum and energy of the innermost stable circular orbit (ISCO) of this spacetime, satisfying $V_m'(r)=V_m''(r)=0$, have the properties \cite{Xamidov:2026kqs}:
\begin{eqnarray}
	r_{ISCO}&=&\frac{6M}{1-B^2M^2}, \label{risco}\\
	L_{ISCO}&=& \frac{2\sqrt{3}M}{1-14 B^2 M^2+B^4 M^4} ,  \label{Lisco}                    \\
	E_{ISCO } &=& \frac{2\sqrt{2}}{3}\sqrt{\frac{(1-B^2M^2)^3}{1-14 B^2 M^2+B^4 M^4}}.\label{Eisco}
\end{eqnarray}
One can show that the radius of the inner circular orbit, $r^-_{cm}$, given in (\ref{rcircmassive}),   is smaller than $r_{ISCO}$, which is another proof that this orbit is unstable, whereas the outer stable circular orbit is always bigger than $r_{ISCO}$. 

\begin{figure}
	\centering
	\includegraphics[draft=false,width=1\linewidth]{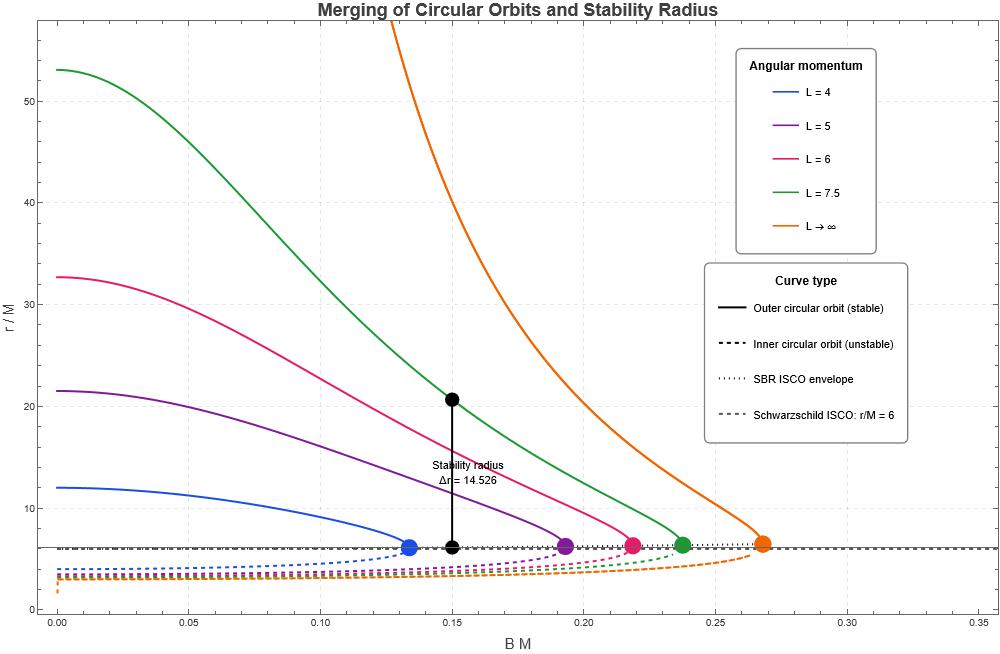}
	\caption{Evolution of inner and outer circular geodesics for increasing $B$ is plotted for $M=1$ and various values of $L$.  We see that for constant $L$, when $B$ increases, these radii get closer to each other and at a certain value $B=B_{crit}$, represented by colorful dots, they merge, corresponding ISCO radius of particular $M$, $L$ values.  The orange curves give the boundary  of circular  geodesics where ultra-relativistic particles orbit the SBR black hole. For $B>B_{max}$, represented by the orange dot, there are no circular orbits  of particles of any type.  For SBR black hole, a vertical line from outer stable radius to $r_{isco}$ gives the coordinate thickness of the region where stable circular orbits can be present, or in some sense determines the distance between inner and outer  edges of the accretion disk around such a black hole. The black vertical bar represents one particular example of this  thickness. }
	\label{fig:circmassmergeB}
\end{figure}

For a given mass and angular momentum, the radius of the inner circular orbit increases and the outer one decreases with increasing $B$. There is a critical value, $B_{crit}$, where these  inner and outer circular orbits merge. This happens when the discriminant in equation (\ref{rcircmassive}) vanishes, given by
\begin{equation}
	B_{\rm crit}(L,M) \;=\; \frac{1}{M}\sqrt{\,7 \;-\; 2\sqrt{3\left(4+\frac{M^{2}}{L^{2}}\right)}\,}.
\end{equation}
This critical value, $B_{crit}$ occurs exactly at the ISCO, $r_{merge}=r_{ISCO}, L_{merge}=L_{ISCO}, E_{merge}=E_{ISCO}$ given by equations  (\ref{risco},\ref{Lisco},\ref{Eisco}). These are shown in Fig.~\ref{fig:circmassmergeB}. Note that there is a maximum value of the magnetic field, $B_{max}$, at which this merge occurs in the limit $L\rightarrow \infty$. Beyond this value, no circular orbits exist. The value at which circular geodesics no longer exist is $B_{max}=(2-\sqrt{3})/M\approx 0.268/M$ with radius $r_{max}=(3+2\sqrt{3})M\approx 6.464M$.   In other words, for $B<B_{crit},$ the effective potential has both a local minimum and a local maximum for sufficiently large $L$, corresponding to stable and unstable orbits $r^\pm_{cm}$, respectively. For $B=B_{crit}$ the potential barrier flattens into a single inflection point. For $B>B_{crit},$ due to stronger magnetic field, the effective potential has no extrema and hence no circular timelike orbit exists for the specified $M$ and $L$ values.  The upper limit for $B_{crit}$, given by $B_{max}$, is the ultrarelativistic limit of this radius where $E,L\rightarrow \infty$ and tangential speed of the particle approaches the speed of light. Namely the inner unstable timelike circular orbit, the outer stable one and, as we will show later, the photon sphere coalesce into a single circular null orbit at $r=r_{max}$. Beyond $B_{max}$, there are no circular orbits of any kind in this spacetime.

\subsection{Closed Orbits and Advance of Perihelion}
We now turn our discussion to bound noncircular equatorial orbits and the associated advance of perihelion. Combining the radial and azimuthal equations in Eq. (\ref{geo1},\ref{geo2}), we obtain the first-order orbit equation
\begin{equation}\label{orbiteq}
	\left( 	\frac{dr}{d\phi}\right)^2=\frac{\dot{r}^2}{\dot{\phi^2}}=\frac{r^4}{L^2}\left[E^2-V_m(r)\right],
\end{equation}
which allows us to express the orbits as $r(\phi)$. The usual $r=1/u$ transformation brings the equation  in a more manageable form, in terms of inverse orbit radius $u(\phi)$, given by
\begin{equation}\label{invorbit}
	\left(\frac{du}{d\phi} \right)^2=\frac{E^2}{L^2}-\left( 1-B^2M^2 -2M u\right)\left(u^2+B^2+\frac{1}{L^2}\right).
\end{equation}
 In order to bring the equation in a form easier to solve, we differentiate the equation (\ref{invorbit}) with respect to $\phi$, which brings the orbit equation into a Binet like form:
\begin{equation}\label{Binet}
	\frac{d^2u}{d\phi^2}+u-\frac{M}{L^2}=3M u^2+M B^2\,\left( 1+M u \right),
\end{equation} 
where the left hand side corresponding to Newtonian orbits giving closed Keplerian   ellipses, whereas the first term in the right hand side is the GR correction due to the Schwarzschild black hole and the last term is the correction due the external magnetic field, $B$. Now we want to determine perihelion precession for SBR metric and how magnetic field affects it. There are various ways to  investigate that. So in the following, we determine the perihelion precession using different methods and compare them to reveal exact form of that term and the effect of the magnetic field on that term.

 \subsubsection{Perihelion Precession by Solving Binet Equation Perturbatively}

Our aim here is to obtain a perturbative solution of Eq. (\ref{Binet}). We choose mass of the black hole, $M$, as the perturbation parameter and expand Eq. (\ref{Binet}) in the powers of $M$, while treating the external magnetic field $B$ as a fixed parameter.  It will turn out that it is better not  to treat  $B$ as a separate perturbative parameter, since it is already coupled to $M$, hence its effects appear naturally at the corresponding orders of the expansion in terms of  $M$. In our calculation we include all terms up to second order in $M$, including the $B$ dependent contribution of order $M^2B^2$. 

It is well known that in the Newtonian limit, i. e., $M=0$, Eq. (\ref{Binet}) reduces to
\begin{equation}
	u_{0,\phi\phi}+u_0-\frac{M}{L^2}=0,
\end{equation}
whose solution is
\begin{equation}
	u_0=A_0(1+e\cos\phi),
\end{equation}
where
\begin{equation}\label{pdef}
	A_0=\frac{M}{L^2}=\frac{1}{p},
	\qquad
	p=\frac{L^2}{M},
\end{equation}
and $p$ is the semi-latus rectum of the Newtonian elliptical orbit,
while $e$ denotes its eccentricity.

However, the SBR Binet equation  (\ref{Binet}) contains the additional constant term
$MB^2$. Since the perturbative expansion is organized in powers of $M$,
this term is of lower order than the explicit $M^2B^2u$ contribution.
It must therefore be incorporated into the background solution rather
than treated as part of the second order perturbation. We therefore
choose the zeroth order equation in the presence of the external
magnetic field as
\begin{equation}
	u_{0,\phi\phi}+u_0-\frac{M}{L^2}-MB^2=0.
\end{equation}

Its solution is
\begin{equation}\label{u0}
	u_0=A(1+e\cos\phi),
\end{equation}
where
\begin{equation} \label{Adef}
	A=A_0+MB^2
	=\frac{M}{L^2}+MB^2.
\end{equation}

We now write the orbit function  as
\begin{equation}
	u=u_0+M u_1+M^2u_2+O(M^3).
	\label{persol}
\end{equation}
Here, $u_1$ and $u_2$ denote the coefficients of the first and
second order corrections, respectively. Since $B$ is kept fixed, the
second order function $u_2$ contains both $B$ independent and
$B^2$ dependent contributions.

At first order, substitution of (\ref{persol}) into the Binet equation (\ref{Binet}), involving  the nonlinear term $3Mu^2$, using the  zeroth-order solution (\ref{u0}),  gives  
\begin{equation}
	u_{1,\phi\phi}+u_1=3\,u_0^2,
\end{equation}
up to order $M$.
We obtain as a solution
\begin{equation}
	u_1=   	3A^2 	\left(
	1+\frac{e^2}{2} 	-\frac{e^2}{6}\cos2\phi 	+e\phi\sin\phi
	\right), 
	\label{u1}
\end{equation}
where the last term is secular and will generate the leading perihelion
advance.

Now using zeroth (\ref{u0}) and first order solution ({\ref{u1}) into the Binet equation (\ref{Binet}), the second order equation becomes
\begin{equation}\label{u2eq}
	u_{2,\phi\phi}+u_2 	=
	6u_0u_1+B^2u_0,
\end{equation}
up to order  $M^2$. The first term on the right hand side results from the second iteration of the nonlinear $3Mu^2$  term, while the second one originates from the explicit $B^2 M^2 u$
term in the Binet equation.

The corresponding particular solution of equation (\ref{u2eq}) is
\begin{align}
	u_{2}
	={}& A\Bigg\{	B^2
	\left(
	1+\frac{e}{2}\phi\sin\phi
	\right)+A^2
	\Bigg[
	18+9e^2 	-3e^2\cos2\phi \nonumber
	 \\&
	+\frac{3}{16}e^3\cos3\phi 	+\left(
	\frac{27}{2}e+\frac{15}{4}e^3
	\right)\phi\sin\phi	 	\label{u2}
\\
	&
-\frac92e\phi^2\cos\phi
	-3e^2\phi\sin2\phi
	\Bigg] \Bigg\} .\nonumber
\end{align}

The perturbative orbit equation up to second order in $M$ is
therefore
\begin{align}
	\label{usol_full}	u(\phi)\simeq{}&
	A(1+e\cos\phi) \\
	& 
	+
	3MA^2
	\left(
	1+\frac{e^2}{2}
	-\frac{e^2}{6}\cos2\phi
	+e\phi\sin\phi
	\right) \nonumber
	\\
	&+
	M^2 A\Bigg\{	B^2
	\left(
	1+\frac{e}{2}\phi\sin\phi
	\right)+ A^2
	\Bigg[
	18+9e^2 \nonumber
	 \\ &
\qquad \qquad	-3e^2\cos2\phi
	+\frac{3}{16}e^3\cos3\phi 
	\nonumber\\
	& \qquad \qquad
	-\frac{9}{2}e\phi^2\cos\phi
	-3e^2\phi\sin2\phi
	\Bigg] \Bigg\}.
	\nonumber
\end{align}

This expression contains all contributions generated by the perturbative
expansion through $O(M^2)$, including the $B^2M^2$ term.
Now we are ready to calculate the perihelion precession angle. Only the secular terms are relevant for extracting the perihelion advance. The oscillatory part of Eq. (\ref{usol_full}) can be written as
\begin{align}
	u_{\rm osc}
	&\simeq{}
	Ae
	\Bigg[
	\cos\phi
	+
	\left(
	3MA+\frac{B^2M^2}{2}
	\right)
	\phi\sin\phi
	\\
	&
	+
	M^2 A^2
	\left(
	\frac{27}{2}
	+\frac{15}{4}e^2
	\right)
	\phi\sin\phi
	-\frac92M^2A^2\phi^2\cos\phi
	\Bigg]. \nonumber
\end{align}

The $\phi^2\cos\phi$ term is precisely the second order term generated
by expanding a shifted oscillation frequency,
\begin{equation}
	\cos[(1-\delta)\phi]
	=
	\cos\phi
	+\delta\phi\sin\phi
	-\frac12\delta^2\phi^2\cos\phi
	+\cdots.
\end{equation}

The coefficient of the last term is proportional to
$(3MA)^2
=9M^2A^2,
$
 up to order $O(M^2)$, confirming the consistency of the secular expansion.
We therefore identify
\begin{equation}
	\delta
	=
	3MA
	+
	\left(
	\frac{27}{2}
	+\frac{15}{4}e^2
	\right)M^2A^2
	+\frac12B^2M^2.
\end{equation}

Using (\ref{Adef}), (\ref{pdef})
and keeping terms through $O(M^2/p^2)$, while
discarding higher order and mixed terms such as $B^2M^3/p$, gives
\begin{equation}
	\delta
	=
	\frac{3M}{p}
	+
	\left(
	\frac{27}{2}
	+\frac{15}{4}e^2
	\right)
	\frac{M^2}{p^2}
	+
	\frac72B^2M^2
	+ O(M^3)
\end{equation}

Since successive periapses occur when the oscillatory phase changes by
$2\pi$,
\begin{equation}
	\Phi=\frac{2\pi}{1-\delta}\approx 2\pi(1+\delta+\delta^2+\ldots)=2\pi+\Delta\phi,
\end{equation}
and therefore
\begin{equation}
	\Delta\phi=2\pi(\delta+\delta^2+\ldots),
\end{equation}
where we keep terms up to $O(M^2)$.
Expanding consistently to second order gives
\begin{align} \label{DeltaphiBinet}
	\Delta\phi
	&=
	\frac{6\pi M}{p}
	+
	\frac{15\pi}{2}(6+e^2)
	\left(\frac{M}{p}\right)^2
	+
	7\pi B^2M^2
	+O(M^3)
\end{align}
 in terms of $p$. Using $p=L^2/M$, the corresponding expression in terms of $L$ can be recovered.  Note that in the regularized version of SBR solution, the only change  in the perihelion precession angle  (\ref{DeltaphiBinet}) is $7\pi B^2M^2\rightarrow 9\pi B^2 M^2$ if we use the range of $\phi$ , $\phi\in [0,2\pi/(1+B^2M^2)].$ The other terms do not change at the displayed perturbative order.  
 
Here, the first term is the leading Schwarzschild perihelion advance, the
second is the second order correction due to mass obtained within the present perturbative parameterization  $p=L^2/
M$ and the chosen definition of $e$, and the third is the leading
contribution of the external magnetic field. First two terms agree with previous works \cite{Hoenselaers,Damour,DEliseo:2010hkz,Schmidt:2011xi,Tucker:2018rgy}.
The magnetic contribution,
$	\Delta\phi_B=7\pi B^2M^2$
has two distinct origins. The contribution
$
6\pi B^2M^2
$
arises from the $MB^2$ term incorporated into the zeroth order
background through
$
A=\frac{M}{L^2}+MB^2,
$
while the remaining
$
\pi B^2M^2
$
comes from the explicit $B^2M^2u$ term in the Binet equation. Thus,
including $MB^2$ in the background is essential for obtaining the
complete magnetic contribution. If this term were omitted from the
zeroth-order solution, only the latter contribution would remain,
leading instead to	$\Delta\phi_B=\pi B^2M^2$ (or $3 \pi B^2 M^2$  for the regularized case, as reported by \cite{Al-Badawi:2026era}). To further verify the coefficient of the $O(B^2)$ contribution, in what follows, we calculate the perihelion advance using both  exact and  numerical approaches and examine whether they agree in the weak field limit.

\subsubsection{Exact and Numerical Calculations of Perihelion Precession}

The perturbative expression obtained above is based on a weak field
expansion and therefore may not remain accurate when the magnetic field
parameter $B$ increases beyond the weak field regime. According to analytical formula (\ref{DeltaphiBinet}), the effect of the external magnetic field is to increase the precession angle in proportion to the leading magnetic contribution, $B^2.$ Note that this analytical result is applicable only for weak field limit, i. e., $M/L\ll 1$, $M^2B^2\ll 1$, hence for large values of $B$, the analytic perturbative formula may no longer be adequate and a numerical approach may be needed.   As an independent test of its range of
validity, we evaluate the perihelion precession directly from the exact
radial orbit equation (\ref{orbiteq}), without using any  approximation or  expansions in either $M$
or $B$. 
Namely
\begin{eqnarray}\label{numericalperi}
	\Delta\phi&=&2\int_{r_{min}}^{r_{max}} \left(\frac{d\phi}{dr}\right)dr-2\pi
	\\&=&2\int_{r_{min}}^{r_{max}} \frac{L\, dr}{r^2 \sqrt{E^2-V_m(r)}}-2\pi, \nonumber
\end{eqnarray} 
where $r_{min}$ and $r_{max}$ are the points where $d r/d \phi=0$, corresponding to perihelion and aphelion points of the orbit.  The exact calculation will help us both verify the weak field formula (\ref{DeltaphiBinet}) and understand the behavior beyond the weak field regime.

Using the transformation $u=1/r$ in (\ref{numericalperi}), defining $k^2=(u_p-u_a)/(u_3-u_a)$ and finally using the transformation $u=u_a+(u_p-u_a)\sin^2\theta$, the exact expression for perihelion precession angle is found as
\begin{equation}\label{exactperih}
\Delta \phi	=\frac{4}{\sqrt{2M(u_3-u_a)}}K(k)-2\pi,
\end{equation}	
where $u_a$ and $u_p$ correspond to two physical roots of the cubic radial polynomial, $u_3$ is the third root where $u_a<u_p<u_3$ and $K(k)=\int_0^{\pi/2}d\theta/\sqrt{1-k^2\sin\theta^2}$ is the complete elliptic integral of the first kind. In order to evaluate the perihelion precession from this exact expression (\ref{exactperih}), for a given closed orbit with corresponding $E$ and $L$ values, and for given black hole parameters $M$ and $B$, we have to determine three roots $u_a,u_p$ and $u_3$ which should be real and satisfy the required order. Then putting these values into (\ref{exactperih}) gives  perihelion advance in radians per radial period. If, however, the orbit is specified by eccentricity $e$ and semi-latus rectum  $p=L^2/M$, we can use the turning point conditions to determine the corresponding energy and roots and proceed with (\ref{exactperih}).  

Using this exact expression (\ref{exactperih}) we have verified the weak field perihelion precession angle expression (\ref{DeltaphiBinet}),  especially the $7\pi$ coefficient in front of the magnetic field term. To do this we  have first determined the three roots of the radial polynomial and then substituted them into  (\ref{exactperih}). We then expand the result in the powers of $M/p$, while keeping the terms including  $M^2$, to obtain weak field perihelion advance. We see that the resulting expression perfectly matches (\ref{DeltaphiBinet}), which provides an independent verification of the weak field result.

\begin{figure}[htbp]
	\centering
	
\subfloat[Comparison of exact and perturbative perihelion precession.\label{fig:precesscomp}]{%
		\centering
		\includegraphics[draft=false,width=0.47\linewidth]{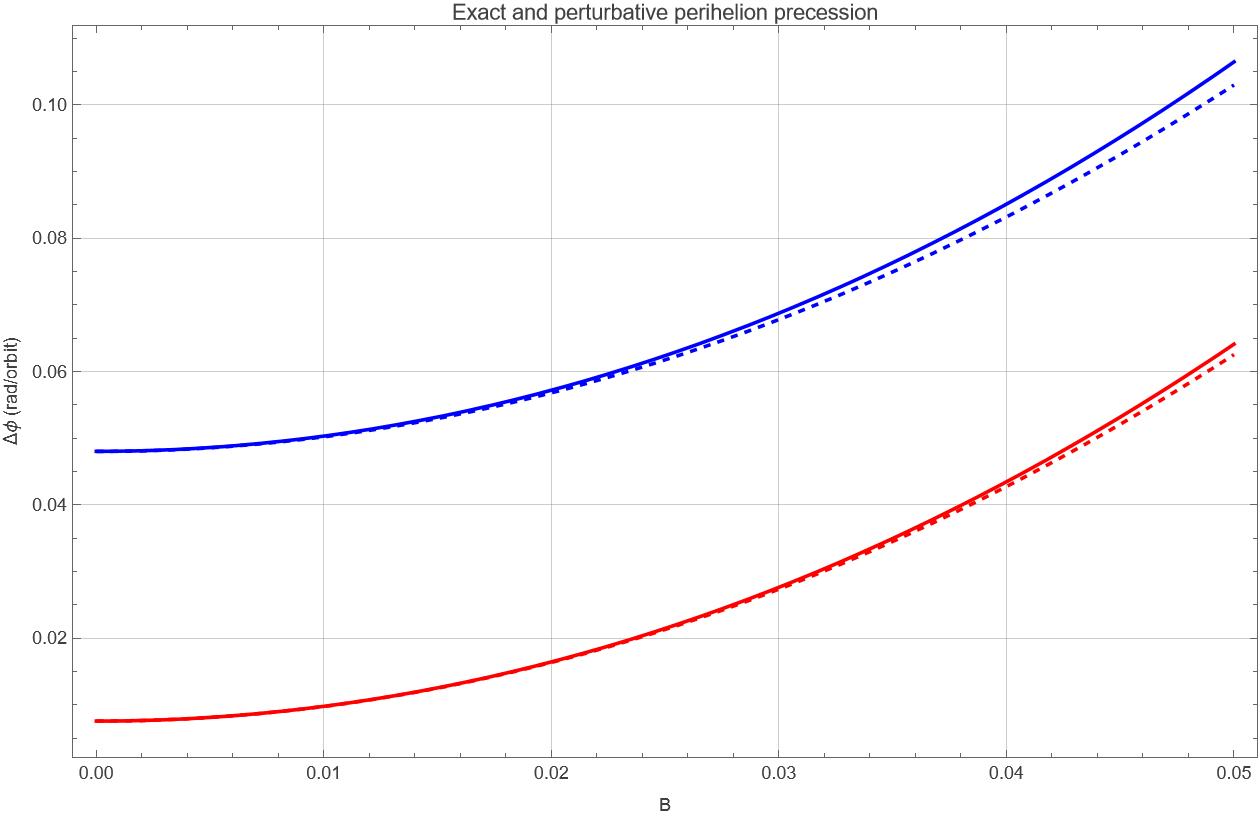}
	}
	\hfill
\subfloat[Difference between the exact and perturbative results.\label{fig:precesscompdif}]{%
		\centering
		\includegraphics[draft=false,width=0.47\linewidth]{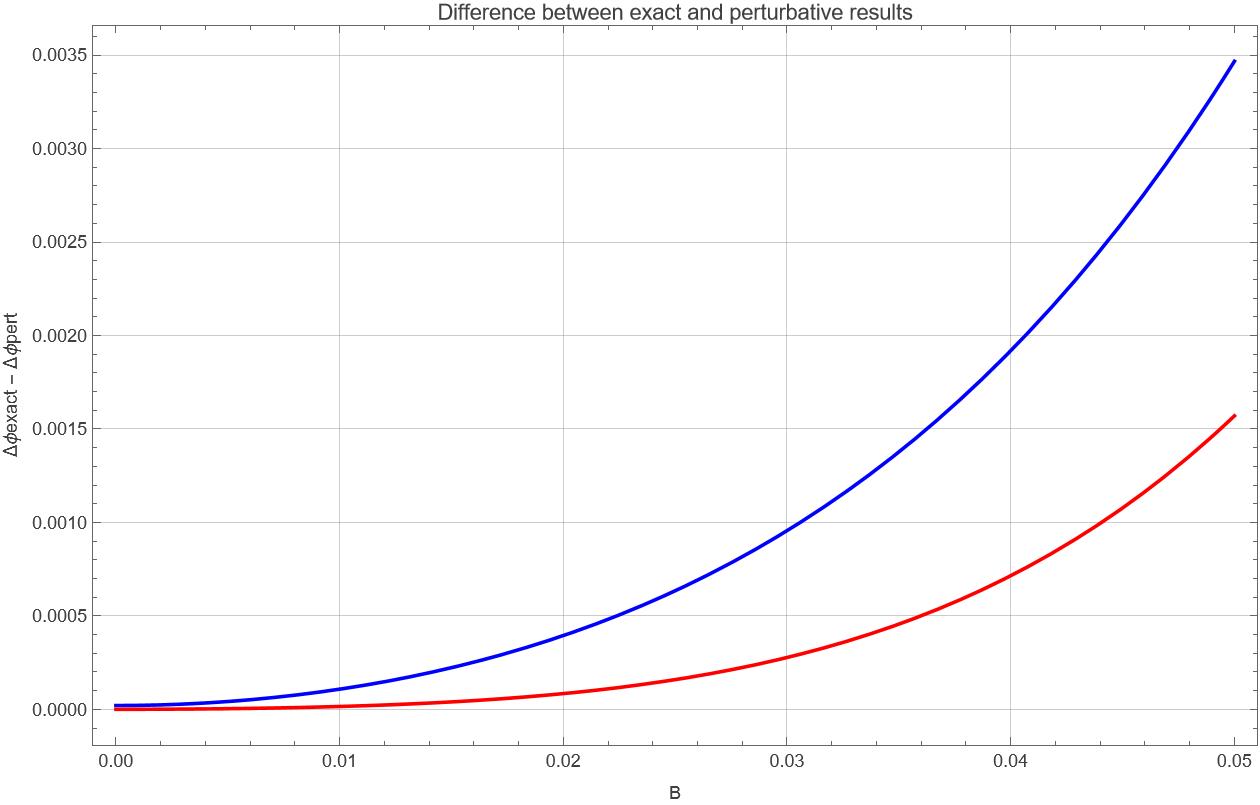}
	}
	
	\caption{Comparison between the perturbative perihelion precession given in
		Eq. (\ref{DeltaphiBinet}) and the exact numerical result calculated from the
		integral in Eq. (\ref{numericalperi}). In panel (a), the perturbative results (dashed curves) agree with the exact numerical results (solid curves) in the weak-field regime and progressively diverge as $B$ increases. The
		calculation is performed for fixed $M=1$, $e=0.05$, and two values of the
		angular momentum, $L=20$ and $L=50$. The energy is calculated for the turning point condition for each  value of free parameter, $B$, when calculating the integral. In panel (b), the corresponding
		difference between the exact and perturbative results is shown for the same
		orbital parameters. The red and blue curves correspond to $L=20$ and
		$L=50$, respectively. In both panels, for each value of $B$, the energy
		and turning points required for the exact numerical integration are
		determined consistently from the fixed values of $M$, $L$, and $e$.}
\end{figure}

We have also evaluated the integral (\ref{numericalperi}) numerically and compared with the analytical result (\ref{DeltaphiBinet}) for two different values of angular momentum, $L=20$, and $L=50$  while keeping the other parameters fixed ($M=1$, $e=0.05$), in Fig. \ref{fig:precesscomp}.  We see that for small $B$, they agree very well. The agreement at small $B$  gives another verification of the weak field result we have found in (\ref{DeltaphiBinet}), especially the coefficient of the term proportional to $B^2$. For larger $B$, however, they start to deviate, with the numerical result (solid curves) giving
a larger value than the perturbative analytic result (dashed curves). The difference between them is plotted in Fig.~\ref{fig:precesscompdif},  confirming that the deviation increases as  $B$ increases, for both cases.  

Note that, for fixed $M$, $L$, and $e$, the family of bound orbits exists only up to a critical magnetic field $B_{\rm crit}(L,M)$.  This fact must also be taken into account when  plotting  the change of  perihelion advance with $B$. Investigating this also provides further insight into the strong field behavior of closed orbits around this black hole. Indeed, in Fig.~\ref{fig:precessstrongB}, we have plotted perihelion advance from 0 to $B_{crit}$ for four different $L$ values representing behavior for small and  large  $L$, while keeping $M=1$ and $e=0.05$. We see that, in all cases, there is a steady increase, but after some value the perihelion advance increases very sharply as $B$ approaches $B_{crit}$. For $B>B_{crit}$ , as we have discussed earlier, the corresponding bound orbits no longer exist.  In Fig.~\ref{fig:precessstrong}, the same results are shown as a function of the normalized magnetic field $B/B_{\rm crit}$. This normalization makes
the common behavior of the perihelion advance for different angular momenta
more apparent.

\begin{figure}[htbp]
	\centering
\subfloat[Perihelion advance as a function of the physical magnetic field $B$.\label{fig:precessstrongB}]{%
	\centering
	\includegraphics[draft=false,width=0.47\linewidth]{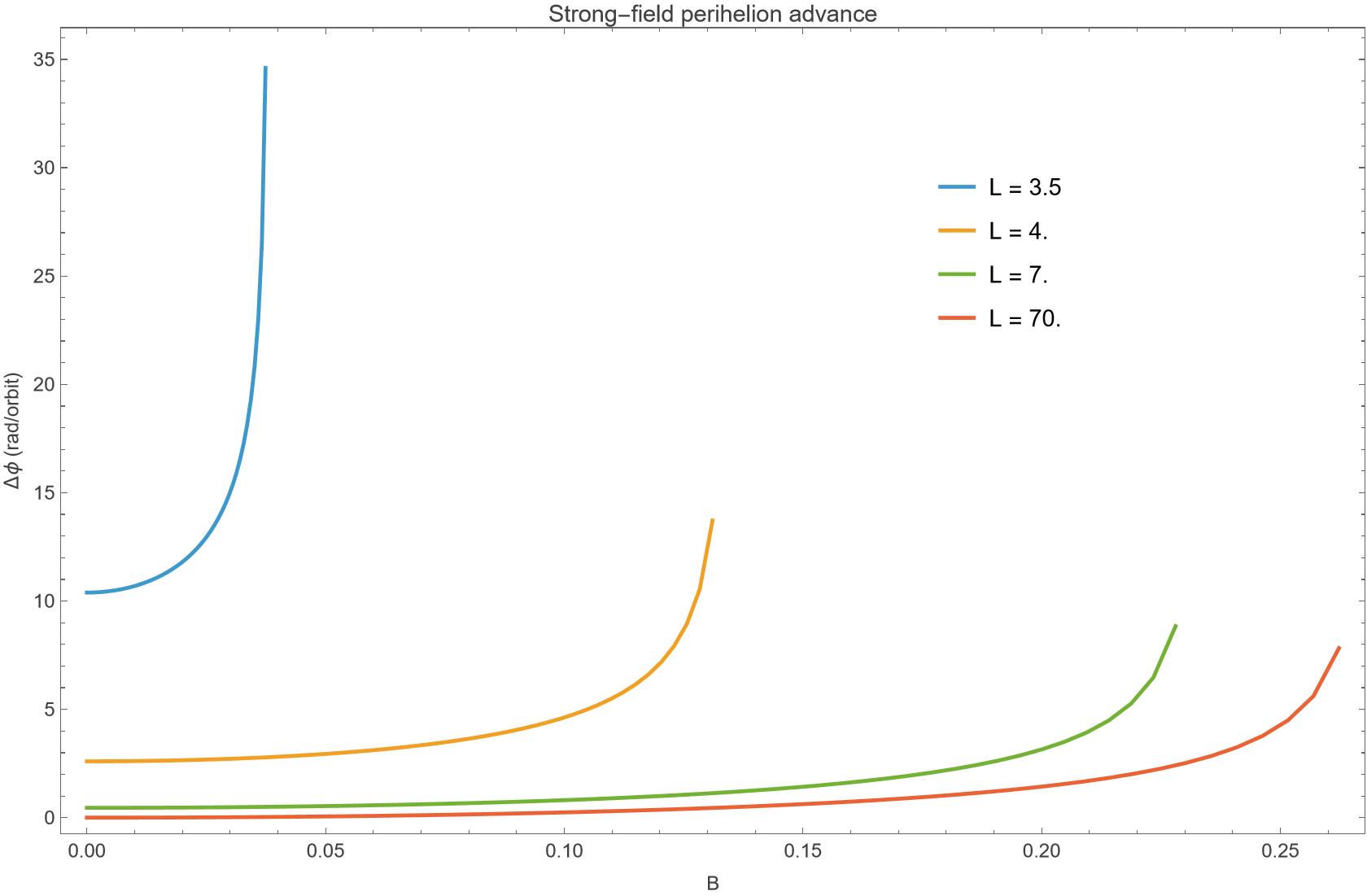}
}	
	\hfill
\subfloat[Perihelion advance as a function of the normalized field $B/B_{\rm crit}$.\label{fig:precessstrong}]{%
	\centering
	\includegraphics[draft=false,width=0.47\linewidth]{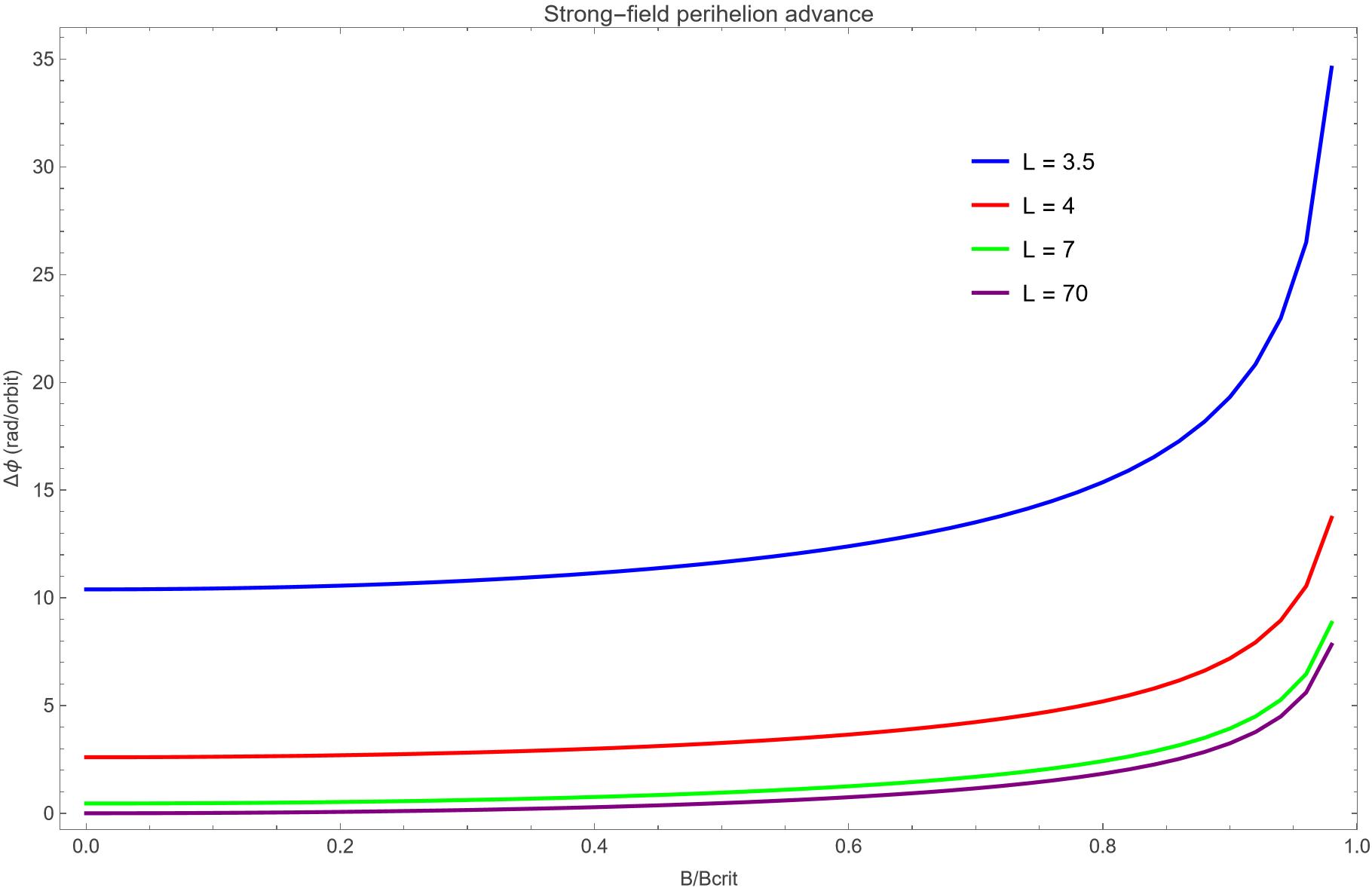}
}	
%
	\caption{Strong-field perihelion advance for different angular momenta.
		The exact perihelion advance is shown for $M=1$ and $e=0.05$, with
		$L=3.5,\,4,\,7,$ and $70$. In panel (a), the perihelion advance is shown
		as a function of the physical magnetic field $B$. Each curve terminates
		at its corresponding critical field $B_{\rm crit}(L,M)$, where the inner
		and outer timelike circular orbits merge. The increase of $B_{\rm crit}$
		with $L$ is evident, with
		$B_{\rm crit}\rightarrow B_{\max}=(2-\sqrt{3})/M$ in the limit
		$L\rightarrow\infty$. In panel (b), the same results are shown as a
		function of the normalized field $B/B_{\rm crit}$. For all angular
		momenta, the perihelion advance increases rapidly as the corresponding
		critical field is approached from below. The large-$L$ curve reflects
		the approach $B_{\rm crit}\rightarrow B_{\max}$ as $L\rightarrow\infty$.}
		\label{fig:precessstrong1}
\end{figure}

As a further independent test, we have considered  the Adkins-McDonnell method \cite{Adkins:2007et}. This method helps to obtain the perihelion advance angle directly by using a central force potential as a perturbation to the Newtonian potential, for a given spacetime geometry, such as Yukawa type or general power law type potentials.  This method has been  especially useful for obtaining perihelion precession of spacetimes  which are solutions of alternative gravity theories \cite{Ozer:2017oik,Ozer:2026fmu}. We have successfully verified the magnetic contribution of our analytical formula (\ref{DeltaphiBinet}) using this method, as another independent verification, although we will not present the details here for clarity. 

\subsubsection{Observational Results}

If we use observational results for perihelion precession, we see that our results are compatible with them. For example, for Mercury, the observed perihelion shift is $43.10\pm 0.10$ arc seconds per century. Here and below, the magnetic-field strength in gauss is converted to
the geometric magnetic-field parameter appearing in the metric through
\[
B_{\rm geom}
=\frac{\sqrt{G}}{c^2}B_{\rm G}
\simeq 2.87\times10^{-25}
\left(\frac{B_{\rm G}}{1\,\mathrm G}\right)\mathrm{cm}^{-1}.
\] 

If we calculate (\ref{DeltaphiBinet}) using a solar magnetic field $B_{\odot}\approx 1\, \mathrm{G}$, it gives a shift $\Delta\phi_B \approx 10^{-38}$ rad, which is much smaller than the observational uncertainty. Hence if the magnetic field of the Sun is modeled as in SBR geometry,  its contribution to the perihelion shift of
the planets is negligible and compatible with observations.   We can also put an upper bound on $B$ by considering observational uncertainty for Mercury, i.e., $0.10$ arcsec per century, then $B_{max}\approx 4.94\times 10^{-11}\, \mathrm{cm}^{-1}\approx 1.71\times 10^{14} \,\mathrm{G}.$ A stronger constraint can be obtained using the supplementary perihelion precession uncertainty from the EPM2011 planetary ephemerides \cite{Pitjeva2013}, 
$0.0030$ arcsec per century,  which gives $B_{\max}\approx8.55\times10^{-12}\,\mathrm{cm}^{-1}
\approx2.97\times10^{13}\,\mathrm{G}.$
These bounds are still much larger
than typical solar magnetic-field strengths, showing that  solar magnetic fields are only  weakly constrained by perihelion-precession observations.

The effect  may be more relevant in regions with strong magnetic fields, such as magnetars.  For a typical magnetar, with $M\approx 1.4 M_{\odot} $, $B\approx 10^{15} \mathrm{G} $, $\Delta\phi_B \approx 7.8\times 10^{-8} $ rad/orbit, which is 30 orders of magnitude larger than the corresponding solar-system
contribution $(10^{-38})$.   Such a large
enhancement suggests that strongly magnetized compact objects may provide
a more promising environment in which to investigate the effects of the
magnetic-field parameter in the SBR geometry. High-precision X-ray timing
observations, such as those performed with NICER \cite{Gendreau2016}, may provide a useful
framework for investigating such effects through the orbital and timing
behavior of strongly magnetized neutron-star systems. 

\section{Lightlike Orbits and deflection of light}

For massless test particles the equations of motion are the same as in equations (\ref{geo1},\ref{geo2}), except that the effective potential in (\ref{effpotm}) should be replaced by 
\begin{equation}
	V_p(r)=\left( 1-B^2M^2 -\frac{2M}{r}\right)\left(1+B^2 r^2\right)\frac{L^2}{r^2}. \label{effpotp}
\end{equation}
We see that as $r\rightarrow \infty$, $V_p$ approaches the finite
value $V_p(\infty)=L^2B^2(1-B^2M^2)$. Thus, photons can reach radial
infinity provided their energy satisfies $E^2>V_p(\infty)$. Therefore, photons can escape to infinity even for $L\neq 0$, which is an important difference from the magnetized black hole solution presented by Ernst \cite{Ernst1976}, where photons cannot escape to infinity, except for $L=0$. 

\subsection{Circular Orbits}

For photons, circular null geodesics  can occur at two distinct radii, given by
\begin{equation}\label{photonradius}
	r_{ph}^{\pm}=\frac{1-B^2 M^2\pm\sqrt{B^4 M^4-14 B^2 M^2+1}}{2 B^2 M}.
\end{equation} 
In the limit $B\rightarrow0$, the smaller-radius branch,
$r_{\rm ph}^{-}$, reduces to $r=3M$, the familiar photon-sphere radius
of the Schwarzschild solution, whereas the larger radius branch,
$r_{\rm ph}^{+}$, approaches infinity. The smaller radius circular photon orbit
is unstable, whereas the larger radius orbit is stable, as follows from
the sign of the second derivative of the effective potential,
$V_p''(r_{\rm ph}^{-})<0$ and $V_p''(r_{\rm ph}^{+})>0$.
Therefore, for small magnetic-field values, $0<B<B_{\max}$, unlike in the Schwarzschild case, two  distinct circular null geodesics exist. At the value for which the radicand in
Eq. (\ref{photonradius}) vanishes, the two photon orbits coalesce into
a single circular photon orbit. Restricting to the physical branch
$B^2M^2<1$, this occurs at the value $B_{\max}=(2-\sqrt{3})/M\approx 0.268/M,$ with the radius $r_{\rm ph,crit}=(3+2\sqrt{3})M\approx6.46M.$
This behavior is shown in Fig. \ref{fig:photonmerge}. For
$B>B_{\max}$, no circular photon orbits exist on this physical branch.

The limiting value, $B_{\max}=(2-\sqrt{3})/M$, also has a natural
interpretation in terms of the ultrarelativistic limit of massive
particles following circular geodesics. As $L\rightarrow\infty$, the critical magnetic field
obtained from the timelike circular-orbit analysis approaches
$B_{\max}$, while the corresponding circular-orbit radius approaches
the radius at which the two circular photon orbits merge. Thus, the
maximum magnetic field parameter obtained from the timelike
circular orbit analysis coincides with the critical field governing
the circular photon orbit structure.

\begin{figure}[htbp]
	\centering
	\includegraphics[draft=false,width=0.75\linewidth]{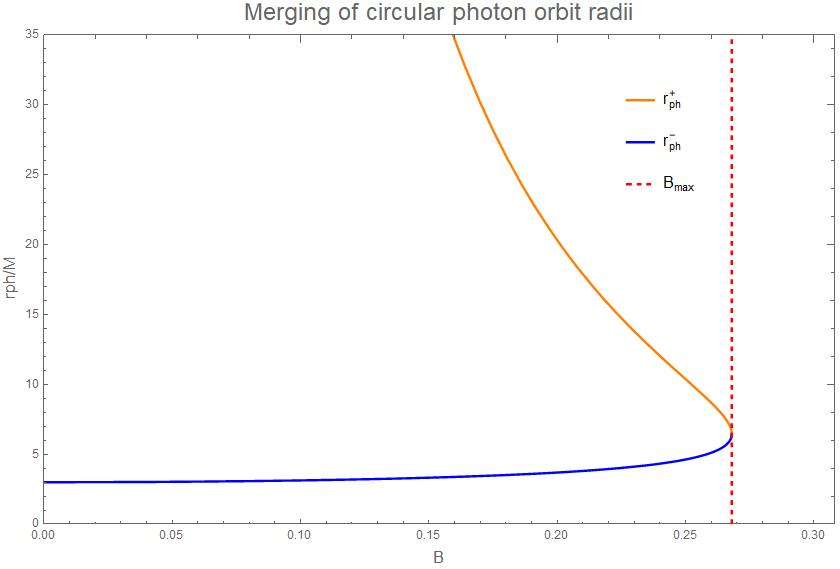}
	\caption{Circular photon-orbit radii as functions of the magnetic-field
		parameter $B$ for $M=1$. The two branches, $r_{\rm ph}^{+}$ and
		$r_{\rm ph}^{-}$, merge at the maximum allowed field
		$B_{\max}=(2-\sqrt{3})/M$. Beyond $B_{\max}$,  there are no circular photon
		orbits in this spacetime.}
	\label{fig:photonmerge}
\end{figure}

 \subsection{Deflection of Light}
 
 \subsubsection{Analytical Perturbative Approach} 
 In this part, we evaluate the orbit equation for photons given by
 \begin{equation}
 	\left(\frac{dr}{d\phi}\right)^2=\frac{\dot{r}}{\dot{\phi}}=\frac{r^4}{L^2}\left[E^2-V_p(r)\right]\label{drdphiph}
 \end{equation}
 where $V_p(r)$ is given in (\ref{effpotp}). Here, our aim is to obtain an analytic result for the deflection angle in the weak field limit. We can use the usual inverse radial coordinate $u=1/r$ which brings the equation (\ref{drdphiph}) into
 \begin{equation}\label{invorbitlight}
 	\left(\frac{du}{d\phi} \right)^2=\frac{E^2}{L^2}-\left( 1-B^2M^2 -2M u\right)\left(u^2+B^2\right).
 \end{equation}
 Taking the derivative with respect to $\phi$ gives the Binet equation for photons in the SBR geometry:
 \begin{equation}\label{Binetlight}
 	\frac{d^2 u}{d\phi^2}+u=3M u^2+MB^2(1+ M u),
 \end{equation}
 where the effects of the black hole mass and magnetic field are clearly seen in the terms on the right hand side. We can again use a perturbative approach to determine the effect  of the black hole mass and external magnetic field on a photon following a straight line trajectory from, say, $-\infty$ to $\infty$, approaching the black hole at the center of the coordinate system to some minimum distance before scattering off. This calculation was not possible for Ernst magnetized black holes, because, unlike SBR geometry, photons cannot escape to infinity in that spacetime.
 
 Here we will obtain  the inverse orbit, $u(\phi)$, in the weak regime, up to order $O(M^2 B^2)$, perturbatively. To do this consistently,  we keep $O(M^2)$ perturbations as well.  Therefore we consider a perturbation of the form
 \begin{equation}\label{uexpansionl}
 	u(\phi)= u_0(\phi)+ M u_1(\phi)+ M^2 u_{2}(\phi).
 \end{equation} 
 In terms of these, the equation (\ref{Binetlight}) gives the following perturbation equations
 \begin{eqnarray}
 	&&u_{0}''+u_0=0, \label{backgroundeq}\\
 	&&u_1''+u_1=3 u_0^2+B^2,\\
 	&&u_2''+u_2=B^2 u_0+6 u_0 u_1,
 \end{eqnarray}	
 where we have replaced $u$ with (\ref{uexpansionl}) on the right hand side of the perturbation equations to keep them on the required order. 
 The total perturbative solution of equation (\ref{Binetlight}) can be found as
 \begin{eqnarray}
 	u(\phi)&=& \frac{\cos{\phi}}{R}+ M\left( B^2+ \frac{3-\cos2\phi}{2R^2} \right) \nonumber \\
 	&+& M^2 \bigg[ \frac{37 \cos\phi+3 \cos{3\phi}+60\, \phi \sin\phi}{16 R^3} \nonumber \\
 	&&\quad  \quad \ +\frac{7 B^2}{2R}\left( \cos\phi+\phi\sin\phi\right)\bigg]. \label{usolnlight}
 \end{eqnarray}
 The solution (\ref{usolnlight}) is obtained such that the path is symmetric under $\phi\rightarrow -\phi$ and $\phi=0$ is the minimum distance between the light trajectory and the source, which can be seen by calculating $u'(0)=0$. When $B=0$, the solution reduces to Schwarzschild case, up to $O(M^2)$ as given, for example, in \cite{Secuk:2019svg}. The solution (\ref{usolnlight}) satisfy the  master equation (\ref{invorbitlight}) if we have  
 \begin{equation}
 	\frac{1}{b^2}=\frac{1}{R^2}+B^2.
 \end{equation}
 Here $R$ is an auxiliary parameter characterizing the zeroth order
 trajectory, while $b=L/E$ is the physical impact parameter.

 Let us now calculate the deflection angle. The setup is such that a  light ray following a path  through $u=0,\phi=-\pi/2-\delta_-$ to $u=0, \phi=\pi/2+\delta_+$ is deflected by the black hole along the way. This path is a straight ray if $M=0$, which can be verified by the solution (\ref{usolnlight}). These small angles of deflection $\delta_{\pm}$ can be calculated by evaluating $u(\phi)$ at these limits, using the smallness of $\delta\pm$. These calculations give that $\delta_+=\delta_-\equiv\delta$, and the  total deflection angle is the sum of these terms as
 \begin{equation}
 	\Delta\phi=2\delta=\frac{4M}{R}+\frac{15\pi M^2}{4 R^2}+2B^2 M R+ \frac{7 \pi B^2 M^2}{2}.
 \end{equation}
 This term involves first and second order contributions to the deflection angle due to the mass of the black hole for light deflection as present in the literature,  and correction terms due to the magnetic field. We can also express this result in terms of the minimum distance, $r_0$, that the light can approach the central object before being deflected by it. This minimum distance can be found from (\ref{usolnlight}) by the relation  $u(0)=1/r_0$, giving
 \begin{equation}
 	\frac{1}{r_{0}}=\frac{1}{R}\left(1+ \frac{M}{R}+M B^2 R+ \frac{ 5M^2}{2R^2}+\frac{7}{2}M^2 B^2\right) .
 \end{equation}
 In order to evaluate deflection angle, we need to express $R$ in terms of $r_0$. Fortunately we only need this at linear order in $M$ since the deflection angle is already linear order in $M$. Therefore we have
 \begin{eqnarray}
 	R&=&r_0 \left(1+\frac{M} {r_0}+M B^2 r_0 \right)+O(M^2),\quad \\ \frac{1}{R}&=&\frac{1}{r_0}\left(1-\frac{M}{r_0}-MB^2 r_0 \right)+O(M^2).
 \end{eqnarray}
 
 With the help of these, we find the desired result as 
 \begin{eqnarray} \label{lightdefana}
 	\Delta\phi&=&\frac{4M}{r_0}+\left(\frac{15\pi}{4}-4\right)\frac{ M^2}{ r_0^2} \\
 	 &&	+ 2B^2 M r_0+ \left(\frac{7 \pi}{2} -2\right)  B^2M^2, \nonumber
 \end{eqnarray}    
 which  is valid  and agrees with the existing literature \cite{Epstein:1980dw,Fischbach:1980su,Richter:1982zz,Edery:2006hm}, including  $O(M^2)$, in the vanishing $B$ limit. The analytical result is valid in the weak-field regime, requiring
 $M/r_0\ll1$ and $Br_0\ll1$ (and hence $BM\ll1$). Note again that, in the regularized SBR solution, the only change to (\ref{lightdefana}) is $7\pi/2 \rightarrow 9\pi/2$ in the relevant term.

To linear order, i. e.,  $O(M)$,  the analytical deflection angle becomes
 \begin{equation}\label{lightapprox}
 	\Delta\phi\approx\frac{4M}{r_0}+2B^2M r_0.
 \end{equation}
 The first term is the usual Schwarzschild term, which decreases with increasing closest approach, $r_0$, whereas the second term is due to the magnetic field and increases with  $r_0$. Therefore, near the black hole the first term dominates and in the far region the second one. If we plot $\Delta \phi$ in this order,  we will have a  $U$ shaped graph with a minimum. The location of the minimum can be calculated by taking derivative of  equation (\ref{lightapprox}) with respect to $r_0$, which gives
 \begin{equation}
 	r_{0,min}=\frac{\sqrt{2}}{B}.
 \end{equation}
 This minimum should be understood as a feature of the truncated
 weak field expression and not as a prediction outside its regime of
 validity. 

\subsubsection{Exact and Numerical  Approach}

In order to verify and explore the domain of  applicability of the analytical result  (\ref{lightdefana}), we need to compare it with the exact or numerical deflection angle result.
For photons, the orbit equation, $r(\phi)$, given in (\ref{drdphiph}),   can be solved  formally   as an integral. 
Using this formal solution, we can determine the deflection angle for photons coming from radial infinity and scattered from the closest approach distance $r_0$, due to the symmetry of the orbit, 
by the following equation 
\begin{eqnarray}\label{numericallight}
	\Delta\phi &=& 2\int_{r_0}^{r=\infty } \left(\frac{d\phi}{dr}\right)dr-\pi \\
	&=& 2 \int_{r_0}^{\infty} \frac{dr}{r^2 \sqrt{\frac{1}{b^2}-\frac{V_p(r)}{L^2}}}-\pi, \nonumber
\end{eqnarray} 
where the effective potential for photons, $V_p(r)$,  is given in Eq. (\ref{effpotp}). In order to evaluate this integral, we need to express $b$ in terms of $r_0$,  which can be calculated from $dr/d\phi=0$, giving
\begin{equation}\label{bsquare}
	\frac{1}{b^2}=\left( \frac{1}{r_0^2}+B^2\right)\left[ \left( 1-B^2 M^2\right)-\frac{2M}{r_0}\right],
\end{equation}
which brings the integral into the form 
\begin{equation} \label{lifgintF}
	\Delta\phi = 2 \int_{r_{0}}^{\infty} \frac{dr}{r^2 \sqrt{ \left( \frac{1}{r_{0}^2} + B^2 \right) \mathcal{F}(r_0) - \left( \frac{1}{r^2} + B^2 \right) \mathcal{F}(r) }}-\pi
\end{equation}
where  
\begin{equation}\label{Fr}
	\mathcal{F}(r)=1-B^2M^2-\frac{2M}{r}.
\end{equation}
The exact integral in Eq.~(\ref{lifgintF}) can  be evaluated
analytically in terms of an incomplete elliptic integral of the first
kind. Introducing the dimensionless variables
\[
x=\frac{r_0}{r},\qquad
q=\frac{M}{r_0},\qquad
\beta=Br_0,
\]
and using (\ref{Fr}), 
Eq.~(\ref{lifgintF}) becomes
\begin{eqnarray}\label{lightintegralx}
&&	\Delta\phi
	=
	2\int_0^1
	\frac{dx}
	{\sqrt{(1-x)}}\times \\
&&	\frac{1}{
			\sqrt{\left[
			1+x-2q(1+x+x^2)-2\beta^2q
			-\beta^2q^2(1+x)
			\right]}}
	-\pi .
	\nonumber
\end{eqnarray}
The factor $(1-x)$ arises from the turning-point condition at
$r=r_0$. Introducing $x=1-t^2$, the integral can be written as
\begin{equation}
	\Delta\phi
	=
	4\int_0^1
	\frac{dt}{\sqrt{A+Ct^2-2qt^4}}-\pi ,
	\label{lightquartic}
\end{equation}
where
\begin{equation}
	A=2-6q-2\beta^2q-2\beta^2q^2,
	\qquad
	C=-1+6q+\beta^2q^2.
\end{equation}
The quartic polynomial can be factorized as
\begin{equation}
	A+Ct^2-2qt^4
	=
	A(1-\alpha t^2)(1-\gamma t^2),
\end{equation}
where
\begin{equation}
	\alpha=
	\frac{-C-\sqrt{C^2+8Aq}}{2A},
	\qquad
	\gamma=
	\frac{-C+\sqrt{C^2+8Aq}}{2A}.
\end{equation}
Using
\[
t=\frac{\sin\theta}{\sqrt{\gamma}},
\]
the integral takes the standard form of an incomplete elliptic
integral of the first kind. Thus, the exact light-deflection angle is
\begin{equation}
		\Delta\phi_{\rm exact}
		=
		\frac{4}{\sqrt{A\gamma}}\,
		F\left(
		\sin^{-1}\sqrt{\gamma}
		\,\middle|\,
		\frac{\alpha}{\gamma}
		\right)-\pi ,
	\label{exactlightelliptic}
\end{equation}
where $F(\varphi|m)$ denotes the incomplete elliptic integral of the
first kind. This expression contains no expansion in either $M$ or
$B$ and provides an exact analytical representation of the
light-deflection angle in terms of the physical minimum approach
distance $r_0$, generalizing well-known  exact deflection angle expressions \cite{Darwin} for the Schwarzschild case to SBR geometry.

As a consistency check, in the limit $M\rightarrow0$ one obtains
$\alpha\rightarrow0$ and $\gamma\rightarrow1/2$, so that
Eq.~(\ref{exactlightelliptic}) gives $\Delta\phi_{\rm exact}=0$, as
expected for a vanishing mass. To recover the perturbative result, we
expand the exact expression, equivalently Eq.~(\ref{lightquartic}), in
the weak-field parameters $M/r_0\ll1$ and $Br_0\ll1$. Keeping terms
through $O(M^2)$ and $O(B^2M^2)$ gives equation (\ref{lightdefana}), confirming the perturbative approach  we have used to obtain the weak field  deflection angle  expression.

\begin{figure}[htbp]
	\centering
\subfloat[Comparison of deflection angle as a function of magnetic field $B$ for weak field analytical and exact numerical results.\label{fig:lightdefcomp}]{%
		\centering
		\includegraphics[draft=false,width=0.47\linewidth]{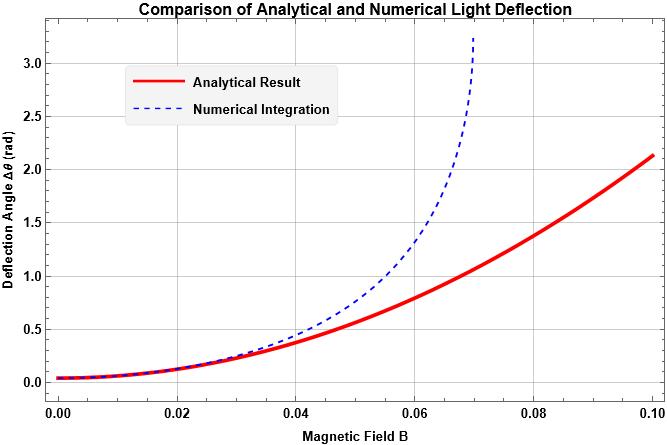}
}
	\hfill
\subfloat[Behavior of deflection angle in the numerical approach as a function of magnetic field $B$ for different mass values, $M=1$, $M=2$ and $M=3$.\label{fig:lightdefmass}]{%
		\centering
		\includegraphics[draft=false,width=0.47\linewidth]{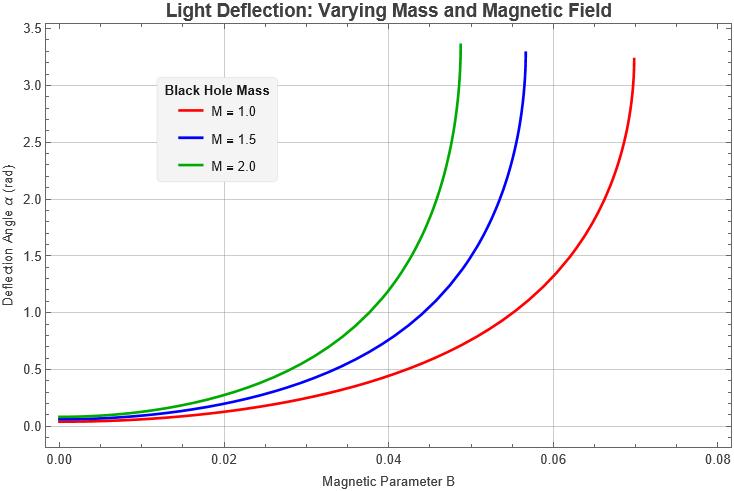}
}
	
	\caption{Change in the  light deflection angle with the magnetic field.  In panel (a) 
	the	weak field analytical light deflection formula given in equation  ($\ref{lightdefana}$) is compared  with the exact numerical value calculated from the integral (\ref{numericallight}).  We see that they both agree in the small $B$ limit, i. e., $B \approx 0.02 $, whereas when $B$ increases, both results diverge. The graph is plotted with $M=1$, $r_0=100$. In the numerical integration, the orbital parameters are calculated dynamically for each $B$.  
		In panel (b), the exact light deflection as a function of $B$ is plotted from the integral (\ref{numericallight}) for three different mass values, $M=1$, $M=2$, and $M=3$ and for $r_0=100$.   We see that, for a given $B$, as $M$ increases, the deflection angle  also increases. Due to the interaction between $B$ and $M$, the bending angle grows  and reaches the limiting value much faster for black holes with increasing $M$. Each curve terminates when the corresponding scattering orbit no longer exists for a given $r_0$, when the radial equation becomes  imaginary.}
		\label{fig:lightdefstrong1}
\end{figure}

We have also numerically evaluated the integral (\ref{lifgintF}) and compared the results  with the analytical formula (\ref{lightdefana}), finding very good agreement in the weak field limit. In Fig. \ref{fig:lightdefcomp}, we see that the analytical formula and exact numerical  results agree very well in the weak field regime. However, when $B$ increases, the results begin to diverge. The analytical result continues to increase smoothly but the numerical result exhibits a logarithmic divergence. This  may mean that for a given $M$ and $r_0$, with increasing $B$ photons can no longer be scattered and the deflection angle increases rapidly so the photons may follow nearly circular orbits and become  confined to them. When $B$ increases, the radius of the stable outer circular photon orbit decreases and near this divergent region, this radius gets closer to $r_0$.

In Fig. \ref{fig:lightdefmass} we explore how deflection angle changes with increasing $B$, using the exact integral (\ref{lifgintF}), for three different mass values. We see that, for fixed $r_0$, the greater the mass, the greater the deflection angle and  curves separate more rapidly and reach the limiting deflection angle. This is due to interaction terms of mass and magnetic field in the relevant formulas. The deflection angle increases and reaches a finite maximum value. Beyond that value, with increasing $B$ photons do not follow the prescribed trajectory of this scattering experiment.  This limiting value can be derived from (\ref{numericallight}) by setting the expression in the square root  to vanish, which gives
\begin{equation}\label{Blim}
	B_{lim}=\sqrt{\frac{r_0-2M}{M r_0 \,(2 r_0+M)}}.
\end{equation}
In other words,  the prescribed scattering trajectory no longer exists for the given configuration, or the photons starting from radial infinity cannot reach $r_0$,  for the given values of the parameters $M$ and $B>B_{lim}$.  

Let us compare our results with the literature. Light deflection in the Ernst spacetime \cite{Ernst1976} is studied in  \cite{Konoplya2006_lens}. In that spacetime, light with a nonzero impact parameter is confined by a far turning point, so, unlike in the SBR spacetime, scattering from infinity is not possible and no asymptotic deflection angle exists. Therefore, to overcome this, bending angle between a source and observer at a finite distance  was numerically calculated in \cite{Konoplya2006_lens}.  The bending angle  increases with $B$ for fixed impact parameter $b$ in the Ernst case, and for fixed closest approach, $r_0$, in the SBR case, though  their dependence  on $B$ may differ.  The strong field behavior is also different, in the Ernst spacetime light is confined, whereas in the SBR spacetime the deflection angle increases with $B$ up to a maximum at $B_{lim}$ given in Equation (\ref{Blim}), beyond which there is no scattering orbit reaching $r_0$ and hence no deflection angle exists.    

\subsubsection{Observational Limits}

Let us now compare our results with solar-system observations. For a
light ray grazing the solar limb, general relativity predicts a
deflection angle of $1.7505$ arcsec. Very-long-baseline interferometry
(VLBI) measurements by Lebach et al.~\cite{Lebach1995} found a
deflection of
$0.9998\pm0.0008$ times the general-relativistic prediction, providing
a direct test of the solar gravitational deflection. Thus, taking the
corresponding observational uncertainty as $
\delta\phi=0.0008\times1.7505
\simeq1.40\times10^{-3}\ {\rm arcsec},
$
we can constrain the magnetic-field contribution.

Using natural units, we take
$M=1475\,\mathrm{m}$ and
$r_0=6.96\times10^{8}\,\mathrm{m}$, corresponding to the solar  mass and radius.
For a representative solar magnetic field
$B_\odot\simeq1\,\mathrm G$, the corresponding geometric-field
parameter is $B_\odot\simeq2.87\times10^{-25}\,\mathrm{cm}^{-1}$. Then, the magnetic contribution obtained from
Eq.~(\ref{lightdefana}) is
$\Delta\phi_B\simeq3.5\times10^{-28}$ arcsec, which is far below the
observational sensitivity. The second-order Schwarzschild contribution
is also small,
$\Delta\phi_{M^2}\simeq7.2\times10^{-6}$ arcsec.

The magnetic contribution in Eq.~(\ref{lightdefana}) contains two terms,
proportional to $2B^2Mr_0$ and $B^2M^2$. Their ratio is $\simeq9.4\times10^4$
so the $2B^2Mr_0$ term provides the dominant magnetic correction.
Using the observational uncertainty above as the maximum allowed
magnetic contribution gives
\[
B_{\rm obs}\simeq5.75\times10^{-13}\,\mathrm{cm}^{-1}
\simeq2.0\times10^{12}\,\mathrm{G}.
\]
This bound is substantially stronger than the corresponding constraint
obtained from Mercury's perihelion precession. Nevertheless, it remains
many orders of magnitude above typical solar magnetic-field strengths,
so the magnetic contribution predicted by the SBR geometry is
negligible for solar-system light-deflection observations.

\section{ Shapiro Time Delay of Radar Echoes}

\subsubsection{Perturbative Approach} 

Here we consider the effects of the external magnetic field on the time delay in a light scattering experiment called the Shapiro time delay, first presented in \cite{Shapiro1964} and observed in \cite{Shapiro1968}. We send a radar signal which passes close to the Sun and reaches a distant planet and is reflected back along the same path. This signal interacts with the gravitational field of the Sun during its journey, causing a time delay compared with the case in which the Sun  is absent. This analysis requires expressing time as a function of the radial coordinate, which can be obtained from  (\ref{geo1},\ref{geo2}) as
\begin{widetext}
\begin{eqnarray}\label{dtdr}
	\frac{dt}{dr}&=&\frac{\dot{t}}{\dot{r}}=\frac{r^2}{\Delta \left(1+B^2r^2 \right) \sqrt{1-V_p/E^2}}
	=\frac{1}{
		\left(1-B^2M^2-\frac{2M}{r}\right)\left(1+B^2r^2 \right)} 
	\left[ 1-\left( 1-B^2M^2-\frac{2M}{r} \right)\left( 1+B^2 r^2 \right)\frac{b^2}{r^2}\right]^{-1/2}. 
\end{eqnarray}
\end{widetext}

Note that the expressions above differ from those of \cite{Konoplya2006_lens} for the corresponding  Schwarzschild-Melvin magnetized black hole solution but reduce to the Schwarzschild case for vanishing $B$. To evaluate this expression, we use the relation (\ref{bsquare}) relating $b$ to $r_0$ in (\ref{dtdr}). 
Then we expand (\ref{dtdr}), keeping the terms up to
$O(M^2)$ and $O(B^2)$, including the mixed terms
$O(MB^2)$ and $O(M^2B^2)$, as
\begin{widetext}
\begin{align}
	\frac{dt}{dr}
	&=
	\frac{1}{\sqrt{r^2-r_0^2}}
	\Bigg[
	r
	+
	M\frac{2r+3r_0}{r+r_0}
	+
	\frac{3M^2}{2r(r+r_0)^2}
	\left(
	4r^2+8rr_0+5r_0^2
	\right) 	-
	\frac{B^2 r}{2}\left(2r^2-r_0^2\right)
\\
	&
-
	\frac{MB^2}{2(r+r_0)}
	\left(
	4r^3+4r^2r_0-2rr_0^2-3r_0^3
	\right)
	-
	\frac{3M^2B^2}
	{4r(r+r_0)^2}
	\left(
	4r^4+8r^3r_0-2r^2r_0^2
	-8rr_0^3-5r_0^4
	\right)
	\Bigg]  \nonumber
	+\mathcal{O}(M^3,B^4). 
\end{align}
\end{widetext}
The time interval experienced by  the signal as it travels from the source at $r_1$ to the minimum distance $r_0$ from the SBR black hole can be calculated as
\begin{widetext}
\begin{align}
	\Delta T(r_0,r_1)
	= &
	\sqrt{r_1^2-r_0^2}
	+
	M\left[
	\sqrt{\frac{r_1-r_0}{r_1+r_0}}
	+
	2\ln\left(
	\frac{\sqrt{r_1^2-r_0^2}+r_1}{r_0}
	\right)
	\right] 	-
	\frac{B^2}{6}
	\sqrt{r_1^2-r_0^2}
	\left(2r_1^2+r_0^2\right)
	\nonumber\\
		&
	+
	\frac{M^2}{4r_0}
	\left[
	\left(\frac{r_1-r_0}{r_1+r_0}\right)^{3/2}
	-9\sqrt{\frac{r_1-r_0}{r_1+r_0}}
	+60\tan^{-1}
	\sqrt{\frac{r_1-r_0}{r_1+r_0}}\,
	\right]
	-
	MB^2
	\sqrt{\frac{r_1-r_0}{r_1+r_0}}
	\left(
	r_1^2+r_0r_1-\frac{r_0^2}{2}
	\right)
	\nonumber\\
	&+
	\frac{3M^2B^2r_0}{8}
	\left[
	\left(\frac{r_1-r_0}{r_1+r_0}\right)^{3/2}
	-\sqrt{\frac{r_1-r_0}{r_1+r_0}}
	-\frac{16
		\sqrt{\frac{r_1-r_0}{r_1+r_0}}
	}{
		1-\frac{r_1-r_0}{r_1+r_0}
	}
	+20\tan^{-1}
	\sqrt{\frac{r_1-r_0}{r_1+r_0}}
	\right]
	+ 	\mathcal{O}\!\left(M^3,B^4\right). \label{DTShapiro}
\end{align}
\end{widetext}

The time-delay expression contains a contribution that is already
present in the $M=0$, $B\neq0$ background. Since our interest is in the
modification of the gravitational Shapiro delay produced by the
magnetic parameter, we define the gravitational time delay relative to
the corresponding magnetic background as
\begin{equation}
	\Delta T_{\rm Shapiro}
	=
	T(M,B)-T(0,B).
\end{equation}
To the order considered here, the purely magnetic $B^2$ contribution
therefore cancels, giving
\begin{align}
	\Delta T_{\rm Shapiro}
	\equiv{}&
	\Delta T(r_0,r_1)
	-\sqrt{r_1^2-r_0^2}
		\nonumber\\&
	+
	\frac{B^2}{6}
	\sqrt{r_1^2-r_0^2}
	\left(2r_1^2+r_0^2\right),
\end{align}
where the second term subtracts the flat-space contribution and the
third term subtracts the purely magnetic background contribution. 
In the limit $B\rightarrow0$, this reduces to the corresponding
Schwarzschild result with second order correction. For $r_1\gg r_0$, the leading terms are
\begin{align}
	\Delta T_{\rm Shapiro}
	\simeq{}&
	M\left[
	1+2\ln\left(\frac{2r_1}{r_0}\right)
	\right]
	+\frac{M^2}{r_0}
	\left(
	\frac{15\pi}{4}-2
	\right)
	\nonumber\\&
	-
	MB^2r_1^2
	-3M^2B^2r_1
	+\mathcal{O}(M^3,B^4).
\end{align}
The last term represents the leading magnetic correction to the
gravitational Shapiro delay.

For a radar signal passing close to the Sun, the total one way 
Shapiro delay is obtained by adding the contributions from the two
radial endpoints. For a trip from a source at radial coordinate $r_1$ to the Sun
and then from the Sun to a reflecting planet at radial coordinate
$r_2$, the total one way Shapiro time delay is given by
\begin{equation}
	\Delta T_{\rm Shapiro}^{\rm total}
	=
	\Delta T_{\rm Shapiro}(r_1,r_0)
	+
	\Delta T_{\rm Shapiro}(r_2,r_0).
\end{equation}

If $r_1,r_2\gg r_0$, the total one way Shapiro delay becomes asymptotically
\begin{align} \label{Shapirotot}
	\Delta T_{\rm Shapiro}^{\rm total}
	\simeq{}&
	2M\left[
1	+
	\ln\left(
	\frac{4r_1r_2}{r_0^2}
	\right) \right]
	+
	\frac{M^2}{r_0}
	\left(
	\frac{15\pi}{2}-4
	\right)
	\nonumber\\
	-
	MB^2&(r_1^2+r_2^2)
	-
	3M^2B^2(r_1+r_2)
	+\cdots.
\end{align}

If we compare the result with that of the Ernst spacetime \cite{Ernst1976}, the only work we are aware of that discusses this phenomenon in that spacetime is \cite{Konoplya2006_lens},  which is  purely numerical, and we see that time delay is growing with $B$, mainly because of the
background magnetic field present even without the black hole. In SBR spacetime, however,  it  decreases with increasing  $B^2$, as shown in the left panel of the Fig.~\ref{fig:timeB}. Therefore, the opposite sign is the main difference between the two spacetimes. Note that all delays are in coordinate time, in both spacetimes, not in the proper time of the observer. If we compare the result (\ref{Shapirotot}) with earlier works in the $B=0$ case \cite{Richter,Teyssandier:2008yx,Ashby}  at  second or higher order in mass, we see that the first order terms agree but the second order terms seem not to agree. This comes from the fact that those works use isotropic coordinates, which affect the second order terms; we expand about $r_0$ whereas they expand about $b$; and they subtract the end point distance as the reference length, whereas we subtract $\sqrt{r_1^2-r_0^2}$ per leg. We have checked that their results and ours agree once transformed into one another.

\subsubsection{Exact Numerical Approach}

Instead of integrating  the equation (\ref{dtdr}) perturbatively, we can numerically integrate  it  to  determine $t=t(r)$. While in the weak field regime the perturbative result (\ref{DTShapiro}) is more practical to use, outside this regime, the numerical approach is necessary. The weak field approximation is usually sufficient for most astrophysical settings. Therefore the high field regime may seem to be a mathematical curiosity, but it may nevertheless be important for extreme cases. In this part, we first show the equivalence of numerical and perturbative results in the weak field regime, confirming the validity of our approximation scheme. Then we discuss behavior outside the limits of the weak field approximation.  

\begin{figure}[htbp]
	\centering
	\includegraphics[draft=false,width=0.8\linewidth]{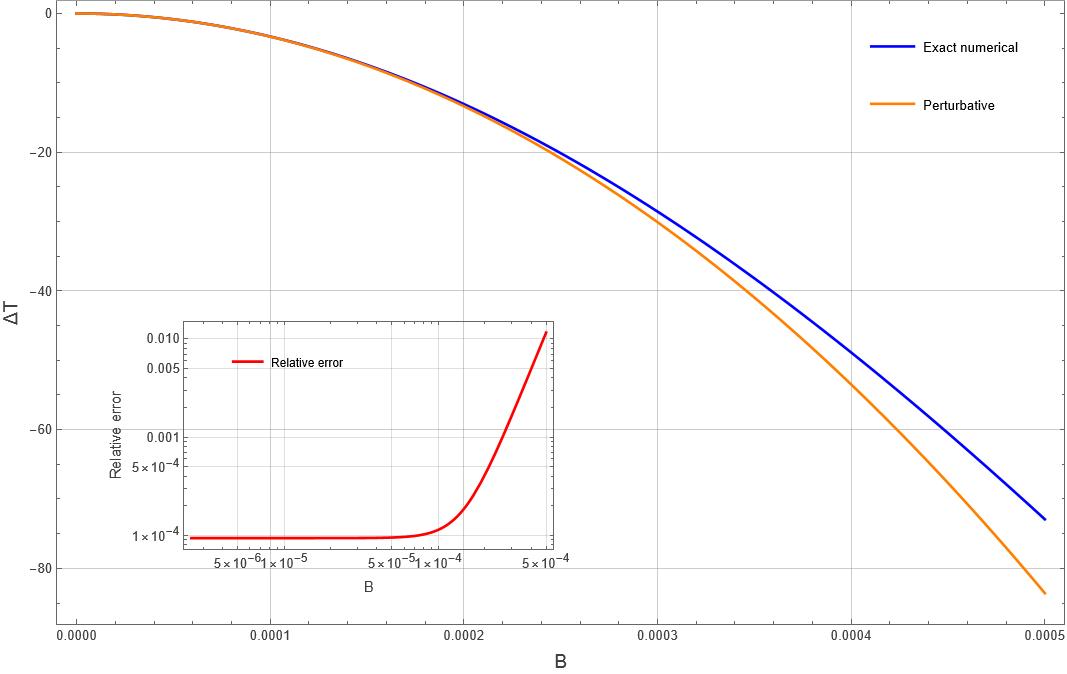}
	\caption{Comparisons of the perturbative and full numerical approaches for $\Delta T$, with $M=1$, $r_0=1000$, $r_1=10000$. The two approaches agree very well in the   weak field regime, while their deviation becomes increasingly obvious as $B$ increases. The relative deviation, shown in the inset figure,  remains approximately constant at small $B$, but increases rapidly beyond a threshold range of $B$, indicating the increasing importance of higher order contributions and the eventual breakdown of the perturbative approximation.  }
	\label{fig:timecomp}
\end{figure}

\begin{figure}[htbp]
	\centering
	\includegraphics[draft=false,width=1\linewidth]{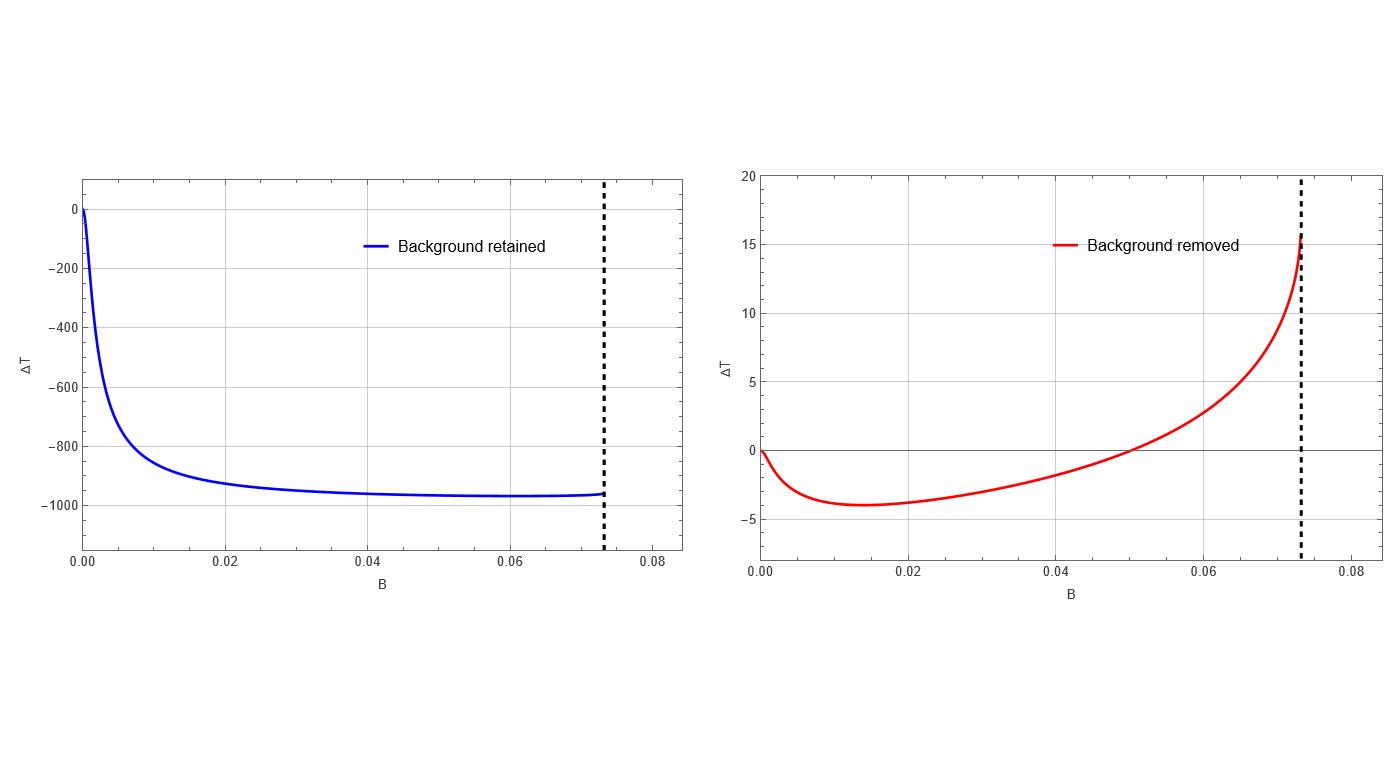}
	\caption{The dependence of $\Delta T$ on $B$, obtained  by numerical integration. The left panel shows full magnetic contribution including background one. The two cases exhibit qualitatively different behavior.  The background term dominates the full behavior and produces a large negative contribution over the entire range. When the background term is removed, the remaining magnetic contribution initially decreases, reaches a minimum, and then rapidly increases as $B$ approaches a threshold value $B_{tr}$. At $B=B_{tr}$, the radicand of the  radial integral vanishes within the integration domain, and the integral becomes nonintegrable. The dashed vertical lines indicate this threshold. Here $M=1$, $r_0=1000,$ $r_1=10000$.}
	\label{fig:timeB}
\end{figure}

Fig.~\ref{fig:timecomp}  shows that in the weak field regime, the perturbative and numerical results agree very well for small $B$, while they begin to deviate as $B$ increases beyond this range. The relative deviation is shown in the inset of this figure. It remains approximately constant at small $B$,  but beyond a threshold value, it increases rapidly, indicating the breakdown of the perturbative approximation as the magnetic field becomes stronger. In Fig. \ref{fig:timeB}, we plot the  time  delay over the full  applicable range of $B$ in the given setting. We see that when the background magnetic contribution is present, the full time delay decreases  monotonically with increasing $B$. In the second panel, where the background contribution is suppressed, the time delay exhibits a $U$ shaped behavior. It initially decreases from its $B=0$ value, reaches a minimum  and then increases further.  The magnitude of the full contribution is much larger than that of the background-suppressed result, implying that the background magnetic term is the dominant term in the magnetic field dependence of the time delay.

\subsubsection{Observational Results}
For a representative solar magnetic field
$B_\odot\simeq1\,{\rm G}$, corresponding to
$B_\odot\simeq2.87\times10^{-25}\,{\rm cm}^{-1}$ in geometric units,
the magnetic contribution is extremely small compared with the
standard Shapiro delay. For an Earth--Saturn configuration, for
example, the magnetic correction is of order
$10^{-27}\,{\rm s}$, whereas the gravitational Shapiro delay is of
order $10^{-4}\,{\rm s}$. Thus the relative magnetic correction is of
order $10^{-23}$ and is far below present radar-echo timing
sensitivity.

We can estimate an upper bound on $B$ using Cassini data \cite{Bertotti:2003rm}. We have two distinct results depending on how the background magnetic field, present even when  $M=0$, is incorporated. If the result is $\Delta T=T(M,B)-T(M,0)$ then we find $B\le  8.52\times 10^{-21}$ $\rm cm^{-1}$, whereas for $\Delta T=T_{Shapiro}(M,B)-T_{Shapiro}(0,B)$ we find  $B\le 2\times    10^{-16}$ $\rm cm^{-1}$. In the former the effect of the background term  of the form $\approx B^2r^3$  is present whereas in the latter it is removed. Note that both upper limits are lower than those obtained  from  perihelion advance and  light deflection measurements, so time delay experiments are more sensitive in this regard.
It should be noted that the above estimates use a constant magnetic
field as a representative value and are therefore only  formal
estimates of the effect of the SBR magnetic parameter. Moreover, the
perturbation expansion employed here requires
$B^2r^2\ll1$ throughout the photon trajectory. Consequently, a
formal upper limit obtained by extrapolating the truncated
$O(B^2)$ expression to larger values of $B$ should not be interpreted
as a rigorous observational bound. A reliable bound in this regime
would require evaluating the exact time-delay integral.
 
\section{Conclusion}

In this paper, we have explored some  solar system weak field phenomena in the Schwarzschild-Bertotti-Robinson spacetime,  known as classical tests of general relativity. These phenomena are important since they are measurable with current technology. Hence they can verify, limit or rule out the theoretical results of theories of modified gravity or new solutions of GR, as in this case.  These and other classical tests are the reasons why Einstein's General Relativity is  considered the theory of gravity, since it passes all of them. The SBR solution, which adds an external magnetic field around a static black hole, has  not been studied in this manner since it is newly introduced. Hence, this paper aims to fill this gap. The first thing we have done in this paper is to obtain analytical results from this solution for three phenomena: perihelion advance, light deflection, and Shapiro time delay, in the weak field regime.

In order to study those phenomena, we first obtained the relevant geodesic equations in this spacetime for  massive and massless test particles. We have obtained  effective potentials  and discussed circular geodesics for both cases. We have verified an important observation that circular geodesics no longer exist if the external magnetic field parameter exceeds $B_{max}=(2-\sqrt{3})/M$  for both massive particles and photons. From this result we can speculate that an increasing magnetic field reduces the stability regions of accretion disks.

 We have successfully obtained weak field analytical expressions for the contribution of the magnetic field for perihelion precession angle, light deflection angle and time delay formula, starting from the relevant geodesic equations and using standard methods. These formulas reveal how the magnetic field in the SBR geometry affects these phenomena.
      
In order to verify these weak field formulas, we have also studied these phenomena numerically using the relevant geodesic equations. In all cases, the weak field formulas are verified by these numerical analyses, confirming the validity of these formulas. Comparisons of these analytical results with the numerical results are presented in graphs for each phenomenon.

We have also studied the strong field effects of these phenomena by integrating relevant geodesic equations numerically over the whole range of the external  magnetic field. With increasing $B$,  we have seen that the  exact numerical results diverge from the weak field analytical ones, indicating that the limits of validity  of the  weak field results have been exceeded.  For perihelion precession, the perihelion angle slowly  increases with increasing $B$, but when $B\rightarrow B_{crit}$ it increases very quickly and blows up. When $B>B_{crit}$, there is no perihelion angle because there are no bound orbits in that region. All of this can be seen in Fig.~\ref{fig:precessstrong1}. For light deflection, we have similar behavior. The deflection angle starts to separate from the weak analytical formula for increasing $B$ and it reaches a maximum value at a limiting  $B_{lim}$ value,  given by equation (\ref{Blim}). There is no deflection angle beyond this limit for a given mass and fixed distance of closest approach. For the time delay, the weak field correction is negative and decreases the delay with increasing $B$. In the numerical results, the full delay decreases monotonically with $B$, while the delay with the background contribution removed first decreases and then increases until the threshold $B_{tr}$, where the integral no longer exist.   
    
The present work can be generalized to more general spacetimes involving  an external magnetic field such as Kerr-Bertotti-Robinson \cite{Podolsky2025} or Kerr-Newman-Bertotti-Robinson spacetimes \cite{Ovcharenko:2026tos} to see the effects of spin or electric charge of the black holes on these observations. These generalizations  may help us to understand the effects stemming from  spin such as frame dragging and interaction of the electric charge of the black hole with the external magnetic field   on the particle and photon motion in the weak field observable domain or in the strong limit. We leave these questions for future work.

\end{document}